# Simulation of gating currents of the Shaker K channel using a Brownian model of the voltage sensor

Luigi Catacuzzeno[*] and Fabio Franciolini

Department of Chemistry, Biology and Biotechnology

University of Perugia, via Elce di Sotto 8, 06123 Perugia - Italy

SHORT TITLE: **A Brownian model of ion channel voltage gating**

[*]**Correspondence to:**

Luigi Catacuzzeno

Department of Chemistry, Biology and Biotechnology

University of Perugia, via Elce di Sotto 8, 06123 Perugia - Italy

**Phone number**: 075-5855755

**Email Address:** luigi.catacuzzeno@unipg.it

# Abstract

The physical mechanism underlying the voltage-dependent gating of K channels is usually addressed theoretically using molecular dynamics simulations. However, besides being computationally very expensive, this approach is presently unable to fully predict the behavior of fundamental variables of channel gating, such as the macroscopic gating current. To fill this gap here we propose a voltage gating model that treats the $S_4$ segment as a Brownian particle moving through a gating channel pore and adjacent internal and external vestibules. In our model charges on the $S_4$ segment are screened by charged residues localized on the other segments of the channel protein, and by ions present in the vestibules, whose dynamics is assessed using a flux conservation equation. The electrostatic voltage spatial profile is consistently assessed by applying the Poisson's equation to all the charges present in the system. The treatment of the $S_4$ segment as a Brownian particle allows to alternatively describe the dynamics of a single $S_4$ segment, using the Langevin's stochastic differential equation, or the behavior of a population of $S_4$ segments, useful to assess the macroscopic gating current, using the Fokker-Planck equation. The proposed model confirms the gating charge transfer hypothesis, with the movement of the $S_4$ segment among 5 different stable positions where the gating charges differently interact with the negatively charged residues on the channel protein. This behavior results in macroscopic gating currents displaying properties quite similar to those experimentally found, including a very fast component followed by a slower one with an evident rising phase at high depolarizing voltages.

# Significance statement

A new modeling approach to explore the mechanism of voltage-dependent gating, based on the description of the channel voltage sensor as a Brownian particle, is presented. The model starts from the structural, geometrical and electrostatic properties of the voltage sensor domain of the channel to predict both the sensor dynamics and the macroscopic gating current. The model qualitatively predicts the main features of Shaker channel voltage-dependent gating and explains them in terms of an electrostatic potential profile originating from all the charges present in the system.

# Introduction

Voltage-gated K (Kv) channels are proteins able to promote a transmembrane flux of K ions in response to a plasma membrane depolarization. They are formed by four subunits surrounding a central permeation pore, each consisting of six transmembrane α-helical segments. The first four of these segments, $S_1$ to $S_4$, form the voltage-sensor domain (VSD), able to change its conformation in response to a membrane depolarization and to allosterically promote the opening of the pore domain formed by the remaining $S_5$ and $S_6$ segments. Mutational and electrophysiological analysis allowed researchers to identify the $S_4$ segment as the main voltage sensing structure within the VSD. In most Kv channels, this structure contains a series of six positive residues (R1-R4, K5, and R6, placed at every third position along the primary structure), four of which (R1 to R4, called gating charges) are thought to couple the membrane voltage changes to the transmembrane movement of the whole $S_4$ segment and to channel opening (Bezanilla, 2000).

Gating charges movement along the electrical gradient can be detected as a small capacitive current, known as gating current, whose time integral at large depolarizations gives a measure of the total charge being transferred across a voltage gradient, that for the most studied Kv channel, the Shaker channel, amounts to 12-14 elementary charges (Schoppa et al., 1992; Seoh et al., 1996; Aggarwal and MacKinnon, 1996). Gating currents of the Shaker channel have been studied in detail in the attempt to obtain information on the mechanism leading to the voltage-sensor conformational changes in response to membrane depolarization. Although the gating current of a single channel cannot be detected directly due to technical limits, fluctuations analysis of the macroscopic gating current is consistent with the presence of 'shot events' of charge movements amounting to about 2.4 elementary charges (Sigg et al., 1994). At the macroscopic level, the (ON) gating currents recorded upon depolarization show complex properties depending on the stimulation protocol used, decaying mono- or bi-exponentially, or showing an initial rising phase. The OFF gating currents recorded on repolarization display likewise composite time courses depending on the prepulse voltage. This complex behavior suggested the presence of multiple stable states for the $S_4$ segment that were interpreted using discrete Markov models (DMM), and assuming the voltage sensor undergoing transitions among discrete, energetically stable conformational states, separated by high energy barriers. (Zagotta et al., 1994; Bezanilla et al., 1994; Stefani et al., 1994; Schoppa and Sigworth, 1998). In the absence of structural information, all these models considered the energetic landscape encountered during state transitions as a parameter, rather than being assessed by applying physical laws. Although this approach has been useful to interpret the experimental data, it is unable to give detailed information on the structure-function relationship and the relevant interactions during the voltage-sensor operation.

The successful application of x-ray crystallography to the Kv1.2 and Kv1.2/Kv2.1 chimera channels provided the first 3D structure of a VSD in its activated conformation (Long et al., 2005, 2007), and indicated that the $S_4$ segment is tilted away from the $S_1$ and $S_2$ helices, to form an extracellular water-accessible vestibule penetrating about 10 Å below the membrane surface. Positive R1-R6 residues on the $S_4$ segment are in part in direct contact with the aqueous solution, in part stabilized by conserved (among Kv channels) external (E183 and E226) and internal (E154, E236, D259) negative clusters, separated by a water inaccessible phenylalanine (F233) located near the midpoint of the membrane. Notably, in the activated state the $S_4$ segment displays for most of its length an $\alpha_{3-10}$-helical conformation, which makes four of the positive charges (R3 to R6) to be separated by 6 Å (instead of 4.5 Å of the α-helix) and aligned towards the interior of the gating pore (Long et al., 2007). Mutagenesis experiments indicate that F233, together with two negatively charged aminoacids of the internal negative cluster, forms a gating charge transfer center (GCTC) that interacts in succession with the R1 to K5 positive residues of the $S_4$ segment during channel gating (Tao et al., 2010). Based on these results it has been proposed that the $S_4$ segment, in addition to the activated and the resting state, can enter three more stable positions corresponding to the different positive gating charges occupying the GCTC (Tao et al., 2010), providing a structural basis for the multiple kinetic states introduced by DMM.

  The new crystallographic data soon inspired a number of different computational studies aimed at predicting the movement of the $S_4$ segment during gating, and the architecture of the channel in the resting and intermediate (closed) states (Yarov-Yarovoy et al., 2006; Pathak etal., 2007; Bjelkmar et al., 2009; Delemotte et al., 2011; Khalili-Araghi et al., 2010; 2012; Schwaiger et al., 2011; Vargas et al., 2011; Jensen et al., 2012). These computational studies all relied on molecular dynamics (MD) simulations, although they differed for applying a force field assessed using either an atom-by-atom approach or more simplified algorithms to reduce the computational effort, such as the coarse-grained approach or the Rosetta methods. Interestingly, all these different approaches, when constrained with some experimental data, reached a consensus regarding the predicted architecture of the resting state of the channel. Namely, the S1-S3 helices retain the conformation assumed in the activated state, while the $S_4$ segment appears rotated and translated inward by about 10 Å, with the R1 residue located extracellularly to F233 along the S2 helix (Vargas et al., 2012; it needs to be stressed, however, that the resting state proposed by these MD simulations contrasts with some experimental results that would instead place the R1 residue in the GCTC , see Tao et al., 2010 and Lin et al., 2011). These computational works has also found evidence for the presence of three stable intermediate states of the voltage sensor (Delemotte et al., 2011), confirming the conclusion of discrete markov models and the more recent GCTC hypothesis, and for the presence of a "focused electrical field" (Islas and Sigworth, 2001; Asamoah et al., 2003; Starace and Bezanilla, 2004) in correspondence of the F233, that allows as many as 3 to 4 charged residues of the $S_4$ segment to completely cross the overall voltage drop (Khalili-Araghi et al., 2010; Delemotte et al., 2011). In one case (Jensen et al., 2012) long (hundreds of microseconds) MD simulations while applying a strong hyperpolarizing pulse (-750 mV) to the open 3D structure show an $S_4$ movement towards a resting state that follows the classical helical screw-sliding helix as the most likely mechanism, with the gating charges not directly exposed to the lipid hydrocarbon, but forming intermediate salt bridges with the external and internal negative clusters.

  Although potentially very useful to understand the physics of the voltage-dependent gating, MD simulations suffer for being very expensive from a computational point of view, due to the need of assessing the movement of hundreds thousands of different atoms every few femtoseconds. This makes it impossible to predict the behavior of experimental variables, such as the macroscopic gating current or the single channel activity, that require either monitoring times much longer than those currently achieved, or the simultaneous assessment of the behavior of a population of homogeneous channels. In order to fill this gap, a number of studies have considered the possibility to use mean field theory models able to start from the available structural data of the VSD and predict macroscopic gating currents, with a computational effort that may be easily achieved by commercially available personal computers (Peyser and Nonner, 2012; Horng et al., 2017). Along this line, here we propose a novel voltage gating model that treats the $S_4$ segment as a Brownian particle moving in one dimension through a water inaccessible gating channel pore and adjacent internal and external water accessible vestibules, delimited by the remaining parts of the VSD (S1 to S3 segments). The treatment of the $S_4$ segment as a Brownian particle allows to alternatively describe the dynamics of a single $S_4$ segment, using the Langevin's stochastic differential equation, or the behavior of a population of $S_4$ segments, using the Fokker-Planck equation to assess the probability density function of finding the particle in the various allowed positions. With these two different approaches, our model can predict the trajectory of a single $S_4$ segment, or alternatively the macroscopic gating currents.

# Results

In our model the geometrical and electrostatic properties of the VSD have been taken from the 3D structure of the Shaker K channel derived by homology modeling, using the available Kv1.2/Kv2.1 chimera structure as a template (Figure 1A; see Supplementary Material). The VSD of the resultant Shaker K channel consisted of a hourglass-shaped geometrical structure made by a short water inaccessible cylindrical gating pore flanked by internal and external water accessible conical vestibules, as inferred from the 3D crystal structure (dashed drawing in Figure 1A). The water inaccessible gating pore was located at the level of the F290 residue, proposed to separate the internal and external vestibules of the VSD. As emphasized in Figure 1C, the $S_4$ segment does not occupy space in either vestibules, since it contributes to form the vestibule walls together with the other transmembrane segments of the voltage sensor domain (Long et al., 2007).

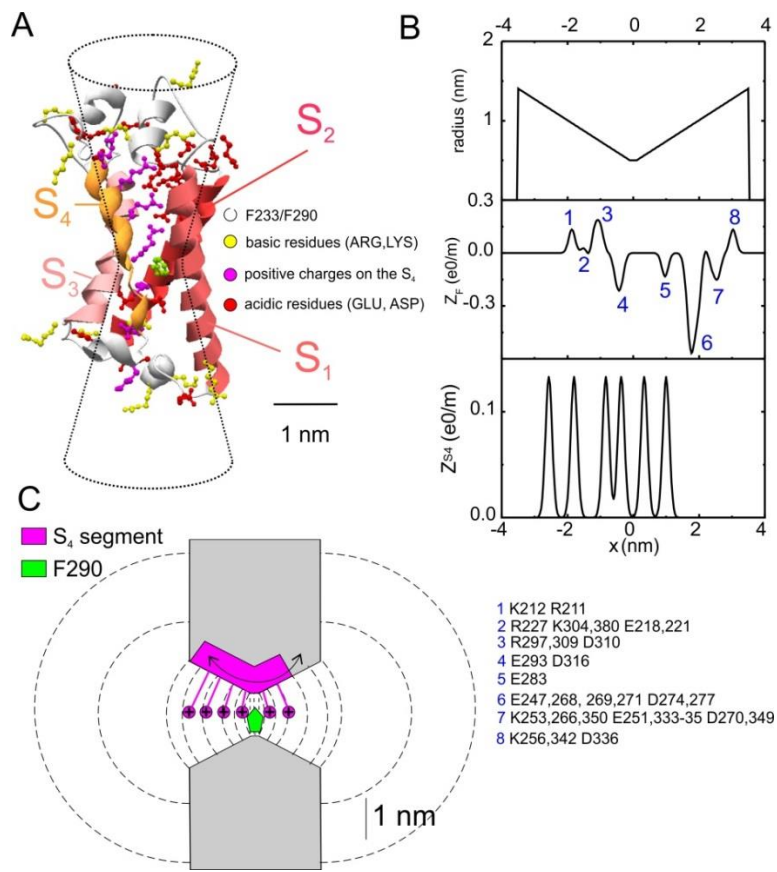

**Figure 1. A)** Model of the Shaker 3D structure obtained by homology modeling using the 2R9R structure as template. Gating charges on the $S_4$ segment are in magenta, while negative and positive residues located in the remaining parts of the VSDs are in red and yellow, respectively. The residue F290 (in Shaker) is in green. The hourglass-shaped drawing superimposed to the Shaker structure represents the geometry used in our model to delimit the gating pore and vestibules. **B)** Profiles of the gating pore radius, the fixed charge density located in the S1-S3 region of the VSD ($Z_F$), and the charge density on the $S_4$ segment ($Z_{S4}$), for the Shaker model structure. For symmetrical reasons, x=0 was assumed to coincide with the center of the gating pore, where the F290 residue is located. **C)** Schematics showing the geometry of the VSD assumed in our model. The $S_4$ segment containing the 6 gating charges was assumed to move perpendicular to the membrane through the gating pore (0.2 nm long) and the extracellular and intracellular vestibules (each 3.4 nm long, and opening with a half angle of 15°). The dashed lines represent some of the surfaces delimiting the volume elements considered in our numerical simulations (see Supplementary data for details).

The model explicitly considers the six positively charged residues on the $S_4$ segment, by letting them contribute to the $S_4$ charge density profile ($Z_{S4}$, Figure 1B). By analogy, the fixed charge density profile ($Z_F$, Figure 1B) was built by considering the position of all the positive and negative charges of the S1-S3 segments of the VSD (marked as red and yellow residues in the 3D structure of Figure 1A). The $S_4$ segment (i.e. its charge density profile) was allowed to move as a rigid body along the gating pore and vestibules by Brownian dynamics, and the electrolyte ions located in the vestibules and surrounding baths were subjected to electro-diffusion governed by a flux conservative equation. Finally, both ions and $S_4$ segment dynamics were driven by an electrostatic potential self-consistently assessed by considering all the charges present in the system using the Poisson's equation (see Supplementary Material for details of the model).

### Dynamics of a single $S_4$ segment

Figure 2A shows exemplificative simulations of the dynamics of the $S_4$ segment, obtained by solving the stochastic differential Langevin's equation at 4 different applied membrane voltages. The time dependent variable $x_{S4}$ plotted in Figure 2A represents the position of the $S_4$ segment, namely the distance of the midpoint between its R2 and R3 residues from the center of the gating pore, where F290 (in Shaker channels) is located. When $x_{S4}=0$, the $S_4$ segment is positioned halfway along its allowed pathway from its furthest intracellular and extracellular positions ($x_{S4}=\pm 1.8$ nm; cf. Figure 1B).

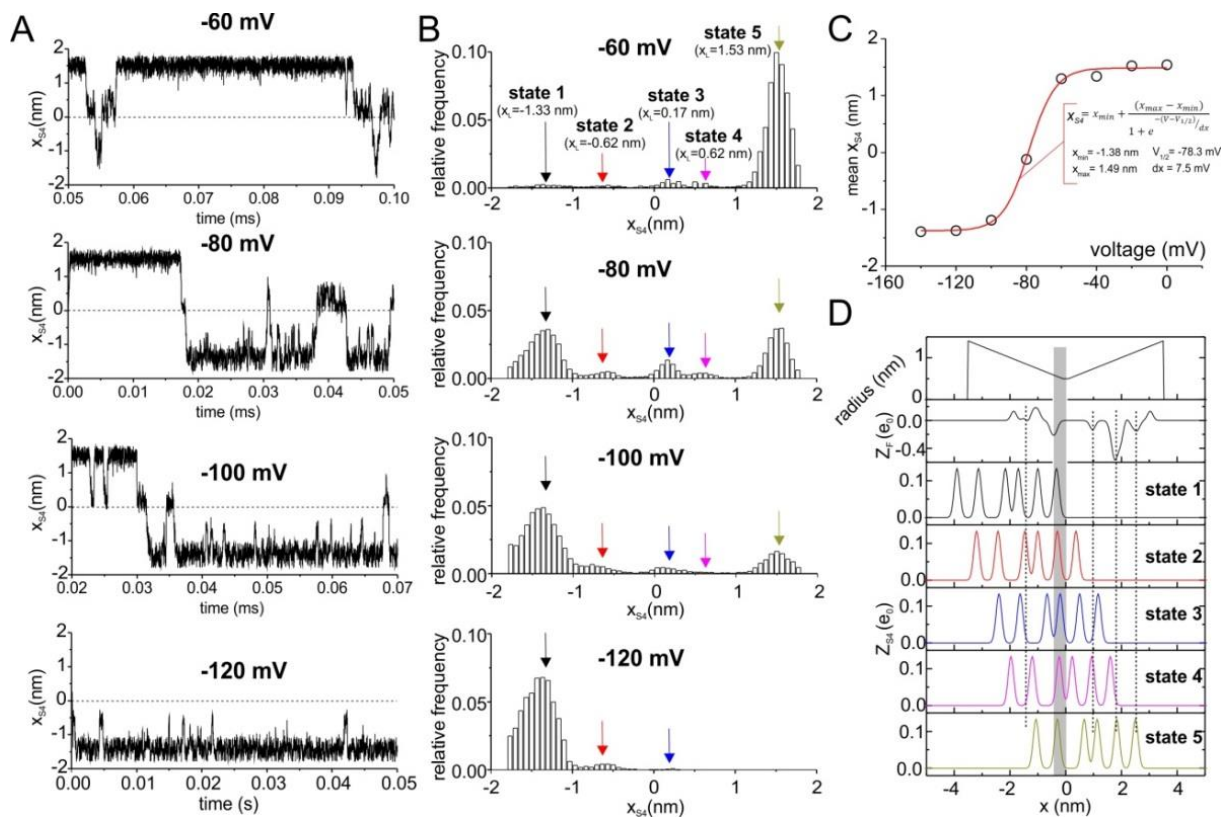

**Figure 2. A)** Exemplificative time courses of the position of the $S_4$ segment ($x_{S4}$), obtained from stochastic simulations at four different applied voltages. **B)** Amplitude histograms of $x_{S4}$, obtained from 100 ms simulations at the indicated voltages. The arrows of different colors, corresponding to the amplitude histogram peaks, represent the 5 positions where the $S_4$ segment spends most of its time. **C)** Plot of the mean $x_{S4}$ as a function of the applied voltage, obtained from 100 ms simulations at the indicated voltages. The red line represents a fit of the model data with a Boltzmann relationship, and the best fit parameters are indicated in the panel. **D)** Plot showing the gating charge profiles for the 5 positions where the $S_4$ segment spends most of its time, corresponding to $x_{S4}=-1.33$, -0.62, 0.17, 0.62 and 1.53 nm. The colors used are the same of the arrows used in panel B to indicate the same states. Also shown are the fixed charge profile ($Z_F$, same of that shown in Figure 1B) and the radius profile of the VSD. The gray region represents the GCTC of Tao et al. (2010), i.e. the region laying between F290 and the first negative peak of fixed charge corresponding to the two closest intracellular negative residues.

Several features are evident from these stochastic simulations. First, the S$_4$ segment tends to assume deep intracellular positions (x$_{S4}$ negative) at very negative voltages (i.e. -100/-120 mV), while the opposite occurs at more depolarized voltages, with the S$_4$ segment mostly residing in the fully activated state at -60 mV. This voltage dependence can be better appreciated in Figure 2C, where the mean position of the S$_4$ segment, assessed from 100 ms long simulations, is plotted as a function of the membrane voltage, and fitted with a Boltzmann relationship (V$_{1/2}$ = -78 mV and dx = 7.5 mV; red line). Second, while moving along its activation pathway, the S$_4$ segment tends to spend most of its time around 5 specific positions, as evident from the 5 clear peaks (indicated by arrows) in the x$_{S4}$ amplitude histograms of Figure 2B. Figure 2D illustrates the position of the S$_4$ charge density profiles (Z$_{S4}$) relative to the fixed charge density profile at the 5 identified preferential positions of the S$_4$ segment. As a first approximation, these 5 positions correspond to the 5 gating charges on the S$_4$ segment, R1-K5, occupying in turn the GCTC (marked by the gray bar in Figure 2D). There are however other electrostatic forces to be considered, that establish charge-to-charge interaction between the gating charges and the fixed countercharges. For example the negative extracellular fixed charge closest to the gating pore (E283 in Shaker), will give a substantial contribution to the stability of states 3 to 5, since its position superimposes with each of the gating charges as they slide through upon activation/deactivation of the S$_4$ segment. Similarly, the two prominent positive intracellular peaks of fixed charge form an energy minimum where R1 to R4 lay when the S$_4$ segment assumes states 1 to 4.

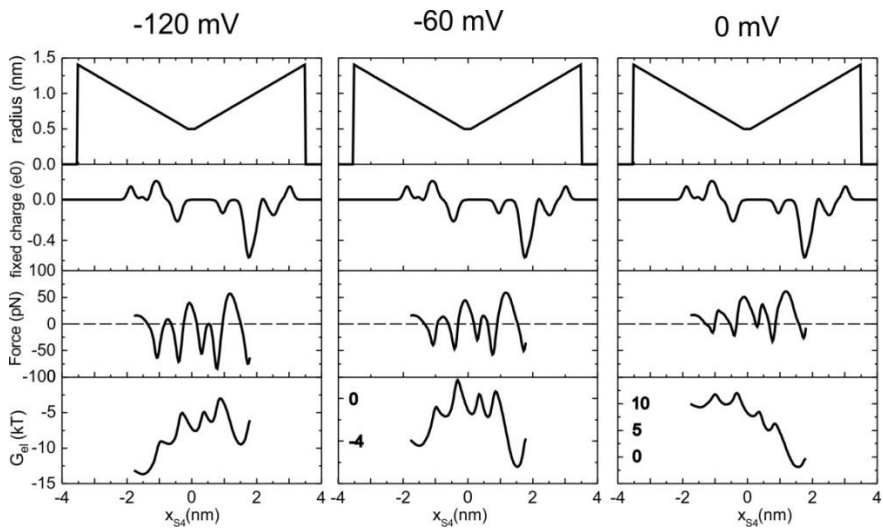

**Figure 3.** Graphs showing the electrical force applied to the S$_4$ segment, assessed using eqn (7) in Supplementary material, and the electrostatic energy associated to the S$_4$ segment, assessed as $G_{el} = \int Z_{S4}\, e_0\, \frac{dV(x)}{dx}\, dx$, as a function of the S$_4$ segment position (x$_{S4}$), at three different applied voltages (indicated). The upper graphs show the radius of the VSD and the fixed charge profile.

To further explore the origin of the 5 substates characterizing the S$_4$ segment, we plotted in Figure 3 the electrostatic energy (G$_{el}$) and the force acting on the S$_4$ segment, for all its possible positions, at three different applied voltages. It is evident from the profiles of G$_{el}$, characterized by 5 electrostatic energy wells, that the electrostatics by itself is able to predict 5 stable positions (states) of the S$_4$ segment. At -60 mV adjacent states are separated by energy barriers of ∼ 2-4 k$_B$T, that must be crossed by the S$_4$ segment during the activation process. At more depolarized or hyperpolarized voltages, the electrostatic energy profile continues to display clear wells, but now the energy barriers separating adjacent wells become even lower than 2 k$_B$T, suggesting an almost continuum motion, rather than a discrete hopping for the S$_4$ segment. In addition, the energy difference among different wells becomes more evident, in a manner favoring strongly states near the deactivated condition at -120 mV, and states near the activated condition at 0 mV. This because now the force acting on the $S_4$ pushes more coherently towards either direction at nearly all positions (i.e. the force is for most of the time negative at -120 and positive at 0 mV).

*Ion concentrations and electrical profiles*
We then looked at the ion concentration and electrical voltage profiles associated with the 5 substates at -120 and 0 mV (Figure 4A). At different positions of the $S_4$ segment correspond very different ion concentration profiles, especially in the extracellular vestibule where the number of gating charges present varies greatly (from 0 to 4) during activation. More specifically, when the $S_4$ segment occupies the most extreme intracellular position (black line, corresponding to state 1, R1 in GCTC), and no gating charge is in the extracellular vestibule, the cation concentration there raises to several hundred millimolars, in order to screen the residing negative fixed charges.

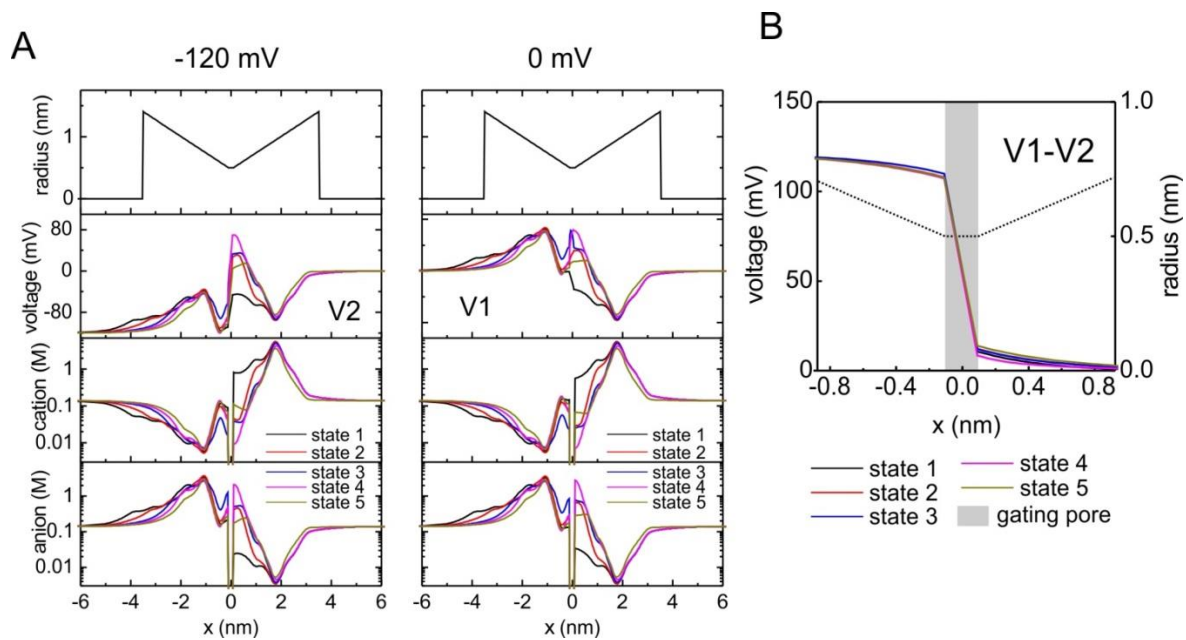

**Figure 4. A)** Graphs showing the voltage and electrolyte ion concentration profiles. The different colors represent simulations performed with the $S_4$ segment positioned at the 5 different stable positions (states 1 to 5). In all cases the simulations were run to equilibrium condition. Left and right panels refer to computations performed at -120 and 0 mV of applied voltage, respectively. **B)** Graph showing the voltage drop profile, assessed as the difference of the voltage profile at two different applied voltages (V2=-120 mV and V1=0 mV, panel A). The different colors represent simulation performed with the $S_4$ segment positioned at the 5 different stable positions (states 1 to 5). The dashed line represents the radius of the accessible zone of the VSD, and the gray region marks the gating pore.

For the same reason anion concentration in the outer vestibule, under these conditions, is very low. As the $S_4$ segment moves towards the extracellular side electrolyte cations tend to leave the extracellular vestibule, while anions concentrate there, creating a negative Debye layer that will screen the gating charges. In the intracellular vestibule the ion concentration changes in relation to the different positions of the $S_4$ segment are much less pronounced, because very close to the gating pore the number of gating charges does not vary as much as seen for the outer vestibule. Unlike the extracellular vestibule, here the gating charges are added or removed much farther away from the gating pore, where volumes per unit length are much larger, and the effect on the Debye screening layer is much less evident. The strong changes in ion concentration profiles in the extracellular vestibule determine in turn strong changes in the electrical voltage profile which very close to the gating pore may change by as much as 100 mV (Figure 4). As a consequence, at 0 mV an $S_4$ segment residing in state 1 will be subjected to a strong force that relocates it to more intermediate positions. Notice in addition that inside the gating pore a highly positive electrical voltage is reached even in the presence of a negatively applied voltage. This repulsive barrier originates from the occurrence that in this region the gating charge may not be effectively screened by electrolyte ions, due to their failure to reach this region. This high voltage barrier however disappears at very negative voltages (i.e., -120 mV) when the $S_4$ segment resides mostly in state 1, since in this condition no gating charges are present inside the gating pore, where screening by electrolyte ions is not possible (cf. Figure 2D). At -120 mV the electrostatic voltage appears in fact to be higher in the extracellular as compared to the intracellular vestibule, while the inverse is true at 0 mV. This explains why at -120 mV the $S_4$ segment is more stable in states where gating charges reside mostly in the intracellular vestibule, while at 0 mV gating charges acquire stability in the extracellular vestibule.

*Focused electrical field across the gating pore*
Several experimental data and modeling results suggest that the structure of the VSD is optimized to allow the focusing of the electrical field produced by the applied voltage within a very narrow region, a condition that would allow the transfer of several gating charges across the entire voltage drop with reduced movement of the $S_4$ segment (Ahern and Horn, 2005; Islas and Sigworth, 2001; Asamoah et al., 2003). This high resistance region is thought to exactly identify with the F290 residue (in Shaker), where the VSD becomes inaccessible to water and ions. In order to test this hypothesis, we looked at the voltage drop profile by assessing the voltage difference profile between -120 and 0 mV. This was done with the $S_4$ segment positioned at the 5 stable states, to verify whether the focusing of the electrical field changes with the position of the $S_4$ segment. As shown in Figure 4B, 90/95% of the voltage drop was indeed concentrated within the gating pore (gray region in Figure 4B), a region 2/2.5 Å long. This is in accordance with experiments indicating a non-zero electrical field across a distance shorter than 4 Å (Ahern and Horn, 2005). In addition our results also indicate that the voltage drop has an essentially identical profile in the 5 different positions (substates) of the $S_4$ segment, suggesting that, at constant applied voltage, during the activation process the electrical field remains essentially the same.

*Simulation of gating currents*
We looked at the gating currents produced by the movement of the $S_4$ segment and associated gating charges among the 5 stable positions. Figure 5A shows a time course of the current that is generated at the boundaries of both the intracellular and the extracellular baths during the movement of the $S_4$ segment ($I_g$, left and right). Also shown are the position of the $S_4$ segment ($x_{S4}$) and the time course of the moving charge, obtained by integrating the gating current. It is evident that the movement of the $S_4$ segment, with consequent charge increase or decrease in the vestibules, induces a remodeling of the ion concentration profile in the vestibules, that in turn causes the appearance of a current in the baths. As expected, identical gating currents are present at the intracellular and extracellular baths, and they appear as a high frequency noise composed of needle-like positive and negative components originating from the random movement of the $S_4$ segment through the gating pore. The time integral of this gating current gives a measure of the charge passing through the gating pore (Figure 5A). As expected for a rigid voltage sensor as the one considered in our model, the time course of the gating charge essentially coincides with the displacement of the $S_4$ segment. Moreover, the charge appears to rest for most of the time at well-defined positions. The corresponding charge amplitude histogram shown in Figure 5B confirms the presence of 5 stable charge levels, each separated from the next by about one electronic charge.

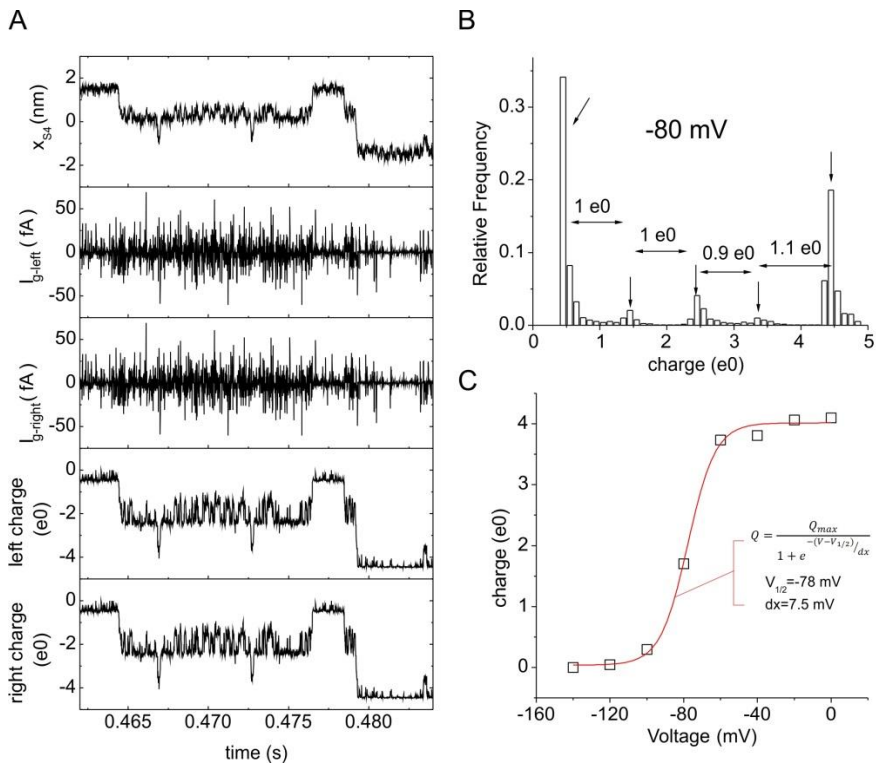

**Figure 5. A)** Graphs showing the position of the $S_4$ segment ($x_{S4}$), the gating current assessed from the changes in net charge of electrolyte ions in the left and right bath ($I_g$(left) and $I_g$(right), respectively), and their integral as a function of time, for a stochastic simulation performed with the full model at -80 mV of applied voltage. **B)** Amplitude histogram of the gating charge assessed from the integral of the left gating charge during 50 ms of simulation at -80. The histogram displays 5 evident peaks separated by about a unitary charge. **C)** Plot of the mean charge vs the applied voltage, obtained by integrating the gating current over 50 ms long simulations at different applied voltages. The solid red line represents the fit of the full model data with a Boltzmann relationship, with best fit parameters indicated.

This is in accordance with the finding that each substate differs structurally from the next by one more charge passing, either way, through the high resistance gating pore. Finally, in Figure 5C we plotted the mean charge assessed from the time integral of the gating currents, as a function of the applied voltage. It is evident that a depolarization results in an increase of the mean charge displacement, up to 4 unitary charges, the maximum number of charges passing from the intracellular to the extracellular vestibule during the activation of the $S_4$ segment. The charge vs voltage (Q-V) relationship could be adequately fitted with a Boltzmann relationship, with parameters mostly identical to those found for the mean $S_4$ segment displacement vs voltage relationship (*cf.* with Figure 2C).

      Although it is not possible to directly record the gating current originating from a single voltage sensor, due to technical limitations, information on the elementary events have been obtained by studying the fluctuation properties of the macroscopic gating current from Shaker channels (Sigg et al., 1994). It was found that the mean current-variance relationship corresponding to the decaying part of the gating current obtained in response to a depolarizing pulse could be interpreted in terms of a simple two state DMM for the voltage sensor, and was in accordance with an underlying shot noise produced by nearly instantaneous transitions of a gating charge amounting to ~2.4 e0. This result suggested that within the gating scheme describing the voltage-dependent kinetics of the Shaker channel there must be a single transition moving a relatively large amount of charge, and placed relatively close to the open state of the channel (Sigg et al., 1994).

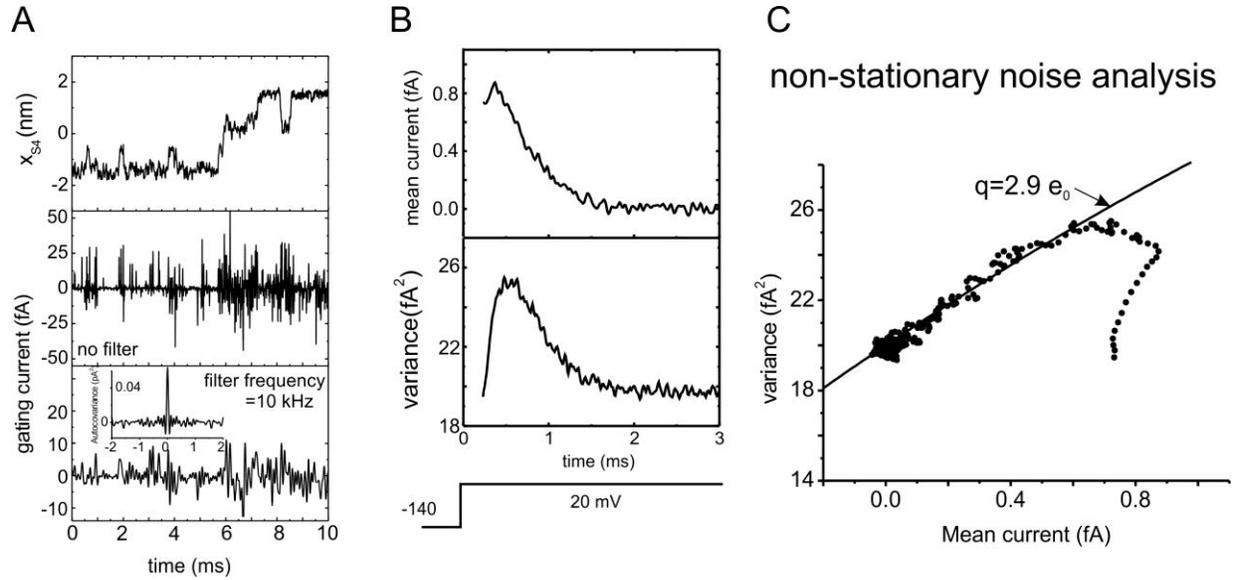

**Figure 6. A)** Plots of the position of the S$_4$ segment (x$_{S4}$), the single-segment gating current, and the same gating current filtered with an algorithm that approximates the effect of an 8-pole Bessel filter with a cutoff frequency of 10 kHz, performed for a 10 ms simulation at an applied voltage of -40 mV. The Inset in the lower plot shows the autocorrelation of the filtered trace, showing a lack of a significant correlation. **B)** Plots of the mean current and variance assessed from 10,000 simulated single-segment gating currents obtained in response to a depolarizing step from -140 to +20mV. Each simulated trace was filtered with an algorithm that approximates the effect of an 8-pole Bessel filter with a cutoff frequency of 10 kHz. **C)** Variance vs mean current plot made with the data shown in panel B. The solid line represents the relationship $\sigma^2 = 2\,B\,q\,i_m - i_m^2$, with q=2.9 e$_0$.

In order to verify whether our model predicts gating current fluctuations having properties similar to those experimentally assessed for Shaker channels, we run 10,000 simulations of 3 ms-long gating current traces originating from a single S$_4$ segment subjected to Brownian motion, in response to a depolarizing step from -140 to +20 mV. After applying a digital 8-pole Bessel filter with a cutoff frequency of 10 kHz to the simulated traces (see Figure 6A), we assessed the mean gating current and the associated variance (Figure 6B), and built the mean current-variance relationship (Figure 6C), which, notably, displayed a shape very similar to that experimentally obtained for Shaker K channels (Sigg et al., 1994). In addition the part of the current-variance relationship corresponding to later times (over 1 ms) from the beginning of the depolarizing pulse could be well fitted with the following relationship valid for a two state DMM: $\sigma^2 = 2\,B\,q\,i_m - i_m^2$ where $\sigma^2$ is the variance, $i_m$ the mean current, B the cutoff frequency of the filter (10 kHz in our case) and q is the elementary gating charge event, which was found to be of 2.9 e0 for our simulation data, a value not much different from 2.4 e0 reported for Shaker K channels. These findings indicate that: 1) our Brownian model qualitatively predicts the fluctuation properties of the gating current experimentally found for Shaker K channels; 2) the interpretation of the mean current-variance relationship with a DMM involving a single transition carrying 2-3 elementary charges does not have a physical basis, since the same relationship is found in a continuum Brownian model where no such transition is evident. It is possible however that, due to the relatively low energy barriers separating the intermediate states (cf. Figure 3), the transition from state S$_2$ to state S$_5$ results in a behavior that, in terms of DMMs, is interpreted as a single step transition associated with the passage of three gating charges through the high resistance gating pore. This may reconcile the interpretation obtained using DMM with our results.

*The fast gating current component*

As already stated, the main advantage of treating the $S_4$ segment as a Brownian particle is that one may alternatively predict the trajectory of a single segment, as shown above, or the behavior of a population of identical $S_4$ segments, by determining the probability density function of finding the segment in the various allowed positions. This second option is particularly attracting as it allows to predict the macroscopic gating currents that have been largely investigated experimentally to monitor the movement of the gating charges.

Figure 7A shows a simulation of the macroscopic gating current obtained in response to a membrane depolarization from -140 to -40 mV (and with the tested assumption that the electrolyte ions equilibrate instantaneously; cf. Supplementary material). Several features of the simulated response have been also observed in experiments. First, within the few microseconds after the beginning of the depolarizing step a very fast gating current component appears, raising instantaneously and then falling very rapidly (*cf.* inset to Figure 7A). This fast component has been experimentally observed using high speed recordings (Sigg et al., 2004). Second, the fast gating current component is followed by a slower component starting with a plateau/rising phase and continuing with a slow decay (Figure 7A, main). The plateau phase disappears at small depolarizations, while it becomes a prominent rising phase for larger depolarizations (cf. Figure 10 and Supplementary Figure 3). All these features of the macroscopic gating currents have been observed experimentally (Bezanilla, 2000).

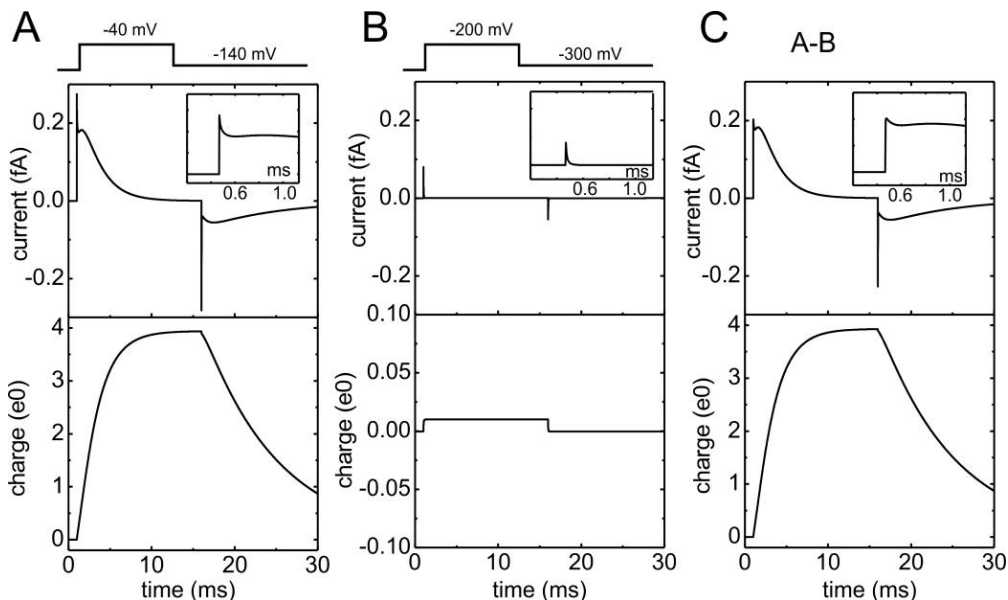

**Figure 7. A)** The upper panel shows a simulated macroscopic gating current evoked by a voltage pulse from -140 to -40 mV (protocol indicated above). The inset is a time expansion of the fast gating current component present at the beginning of the depolarization. The lower panel represents the time integral of the gating current, expressed in units of unitary charges (e0). **B)** Simulated gating currents evoked using a voltage pulse from -300 to -200 mV, to verify whether a fast component of the gating current can be evoked outside the voltage range for $S_4$ segment movement among different states. **C)** Traces obtained from the subtraction of the time course in B from those shown in A, to simulate a leak subtraction as performed in experiments.

In real experiments gating currents are isolated from other types of (linear) capacitive components by standard subtracting protocols (i.e., the currents obtained in response to a depolarizing pulse in a voltage range where the response is no longer voltage-dependent are subtracted from the gating current recorded in the voltage range activating the gating structures). Following this experimental procedure, we simulated the response to a 100 mV depolarization applied from a holding voltage of -300 mV, well outside the activation range of the voltage sensor. As shown in Figure 7B this voltage step evoked only very fast currents resembling the fast component of the gating current shown in panel A. However, the subtraction procedure, shown in Figure 7C, did not completely eliminate the fast component from the macroscopic gating current, indicating that it is not a fully linear component, but a specific feature of the gating current that originates in part from the movement of the gating charges along the activation pathway.

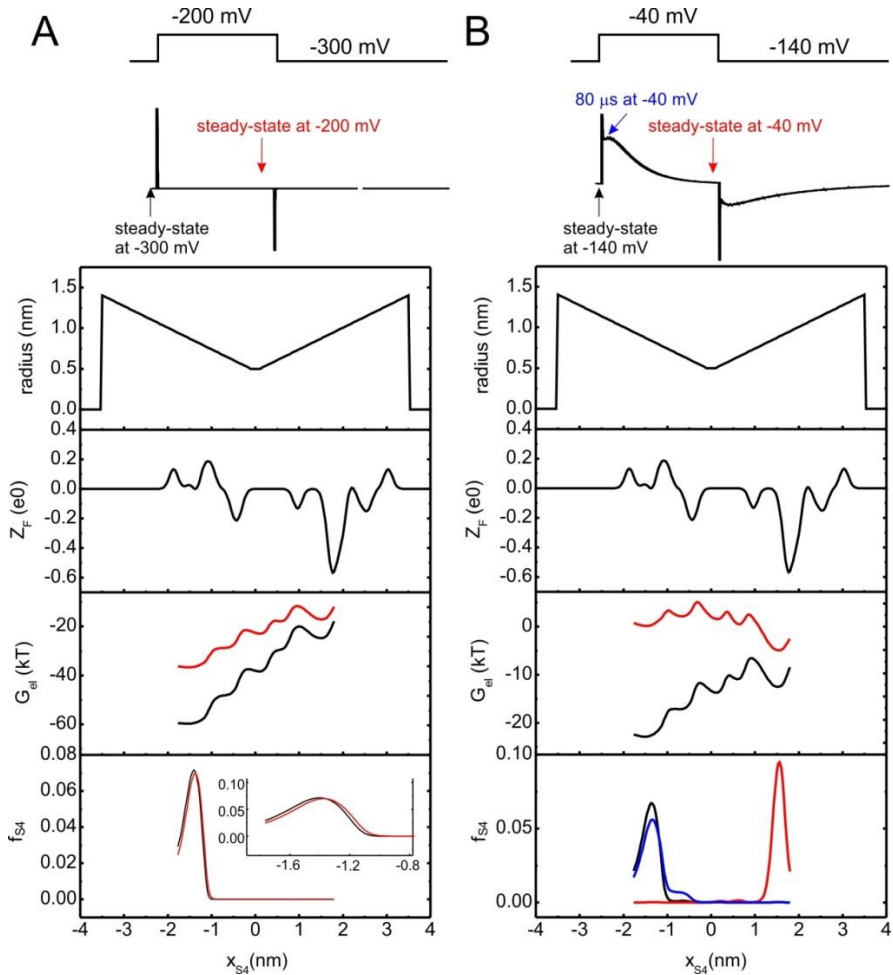

**Figure 8. A)** Profiles of the radius of the VSD, the fixed charge ($Z_F$), the electrostatic energy ($G_{el}$), and probability density function of the $S_4$ segment at the various allowed positions ($f_{S4}$). The black and red lines refer to the times corresponding to the arrows in the voltage protocol shown above, that is when the system is at equilibrium at -300 mV and at -200 mV of applied voltage. The inset in the bottom plot is an expansion of the $f_{S4}$ to evidence the slight different distribution of the S4 segments at the two different voltages. **B)** Same profiles shown in panel A, but during the application of a voltage pulse from -140 to -40mV, just before the depolarization, 80 μs and 15 ms after the beginning of the depolarization.

To understand the origin of the fast gating current component, in Figure 8 we plotted the electrostatic energy ($G_{el}$) and probability density function of the $S_4$ segment position ($f_{S4}$), for the two voltage protocols used in Figure 7. More specifically, in panel A we plotted the above mentioned profiles at -300 and -200 mV at steady-state, in order to understand the origin of the fast gating current component still present at voltages outside the channel activation voltage range, while in panel B we plotted the same profiles at -140 and -40 mV at steady-state, as above (with reference to -40 mV, when both fast and slow current components have been subsided), but also 80 μs after the beginning of the depolarization to -40 mV, when most of the fast gating current component is completed (cf. Figure 7). The comparison of the gating charge distribution at -200 and -300 mV (Figure 8A) shows that the fast component still present outside the voltage range of channel activation results from a slight redistribution of the $S_4$ segments within the energy well of state 1. This because, at both -300 and -200 mV the $S_4$ segments are strongly pushed and confined into state 1, yet at the more negative voltage (-300 mV) the force exerted by the field on the $S_4$ segments is stronger and thus shifts the distribution of the positively charged $S_4$ segments more towards the intracellular boundary. This result suggests that the fast gating current component modeled in a voltage range outside channel activation is due to a charge redistribution within the energy well associated with the extreme stable state. Comparable, but opposite results are obtained at strongly positive voltages (data not shown).

The situation appears quite different when looking at the $S_4$ segment distribution during the pulse from -140 to -40 mV. At -140 mV the $S_4$ segments are mostly positioned in state 1 (black line in Figure 8B). Conversely, after 15 ms at -40 mV, the $S_4$ segments are moved at the most extracellular positions, largely coinciding with the state 5 (red line in Figure 8B). Interestingly, after just 80 μs spent at -40 mV the probability density function of the $S_4$ segment has already changed its shape compared to -140 mV, with a sensible reduction at positions corresponding to state 1, and an increase at positions corresponding to state 2. This result suggests that 80 μs is a time already sufficient to promote the passage of a measurable fraction of $S_4$ segments from state 1 to state 2. This transition would in turn cause the disappearance of a unitary gating charge per $S_4$ segment from the intracellular vestibule, and its appearance in the extracellular vestibule, a movement that will generate a gating current. Thus the fast gating component has two main contributing factors, a linear component due to the re-equilibration of the $S_4$ segments in the energy well occupied at the moment of the depolarization, and a second non-linear component originating from the transition of a fraction of $S_4$ segments from state 1 to state 2.

***The slow gating current component***

We then looked at the slower component of the gating current, i.e. the plateau/rising phase followed by the decaying phase. Figure 9 shows the time course of the gating current obtained in response to a membrane depolarization from -140 to -20 mV, together with quantities that may help to understand their dynamics, namely, the mean position and velocity of the $S_4$ segment, the force acting on it, and the portion of the gating charge residing inside the gating pore, where, as we have seen, most of the electrical field is concentrated (cf. Figure 4B). From the time course of this last quantity it appears that at -140 mV a very small charge is present inside the gating pore (about 0.05 $e_0$, Figure 9Ae), as the $S_4$ segments mostly reside in state 1, with the R1 center of charge positioned in the GCTC. As soon as the VSDs are depolarized, the average gating charge inside the gating pore begins to increase, reaching a maximum of ~0.2 $e_0$ at ~1 ms from the beginning of the depolarization (Figure 9Ae), and then decreases to a lower value (~0.16 $e_0$, Figure 9Ae) that is maintained for the rest of the depolarization. This behavior is mainly due to the coherent movement, upon depolarization, of the $S_4$ segments towards the extracellular vestibule.

Taking into consideration the Fokker-Planck equation (eqn. 10 in Supplementary material), we can rather think of the movement of the $S_4$ segment as being composed of the sum of a voltage driven (drift) term ($-\frac{\partial(\mu(x) f_{S4}(x,t))}{\partial x}$) and a diffusive term ($\frac{k_B T}{\gamma} \frac{\partial^2 f_{S4}(x,t)}{\partial x^2}$). If the drift term is sufficiently high as compared to the diffusive term, the $S_4$ segments will all move at a very similar velocity ($\mu(x)$) towards the extracellular vestibule. Given that all the $S_4$ segments at the beginning have similar positions (cf. Figure 2B, C), we should initially observe an oscillating behavior of the force and of the gating current. In order to confirm this interpretation, we performed a simulation in which the charge density along the S4 segment was made constant, so that the amount of gating charge inside the gating pore would remain constant during the whole movement of the $S_4$ segment. As expected, in this case the gating current was totally devoid of the rising phase (Figure 9B, red).

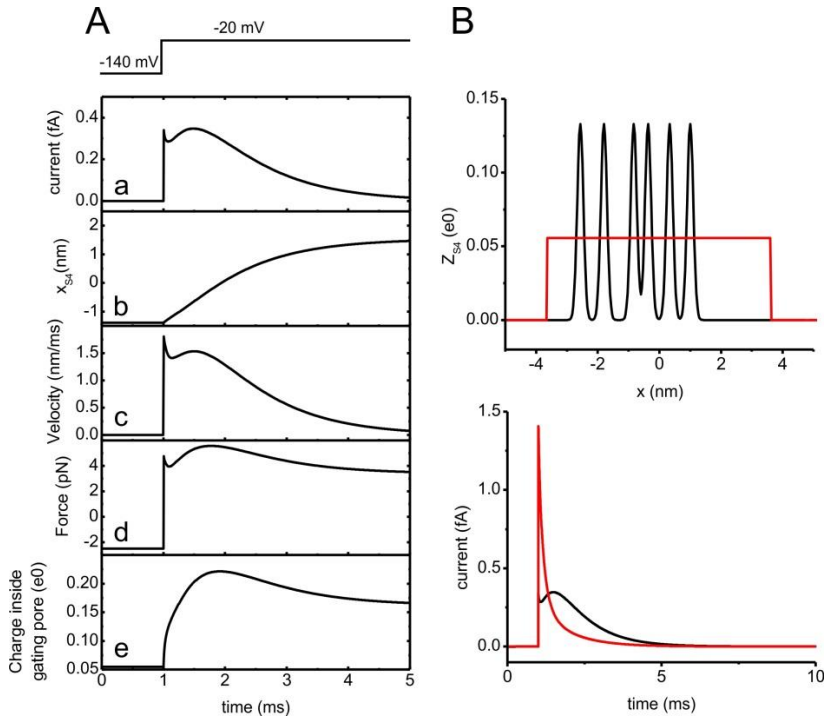

**Figure 9. A)** Time courses of (from top to bottom): the gating current, the mean position of the $S_4$ segment ($x_{S4}$), obtained as $\overline{x_{S4}} = \int f_{S4}\, x_{S4}\, dx$, the mean velocity of $S_4$ segment, assessed as the time derivative of its mean position, the force acting on the $S_4$ segment, and the gating charge residing inside the gating pore. The VSD was subjected to a depolarization from -140 mV to -20 mV (voltage protocol indicated above the plots). **B)** $S_4$ segment charge profile (upper graph) and gating currents obtained in response to a depolarization from -140 to -20 mV (lower graph), for our reduced model (black lines) and for a model in which the gating charge density along the $S_4$ segment was made homogeneous for all the region traversing the gating pore during activation (red lines). Notice that the rising phase of the gating current disappears when considering a homogeneous charge density along the $S_4$ segment.

*Comparing modeled and experimental Q-V relationships*

Finally, we simulated a family of macroscopic gating currents evoked by depolarizations from -120 to -20 mV in 10 mV steps, from a holding voltage of -140 mV, and assessed the Q-V relationship from the time integral of the currents at the various applied voltages (Figure 10A and B). Notably, when the kinetics of the simulated ON and OFF macroscopic gating currents were analyzed in detail, we found a surprising qualitative agreement with the kinetic features found for experimental macroscopic gating currents of Shaker channels (Supplementary Figure 3). Figure 10B shows that the gating charge assessed from the gating currents raises with the applied voltage up to values close to ~4 $e_0$, thus predicting a single channel gating charge of ~16 $e_0$, slightly higher than the 12-14 $e_0$ experimentally found. It also shows that the voltage range where the voltage sensor is predicted to move (-100 mV/-50 mV) is about 20/30 mV more hyperpolarized than that found experimentally for the macroscopic gating charge of Shaker channels. Notable to this regard is that our modeling results are much more similar to the Shaker mutants where the movement of the voltage sensor has been uncoupled from the opening of the channel (Blunck and Balutan, 2012), which is the setting of our simulations where the voltage sensor moves independently on the pore domain. This feature of the model could also explain the discrepancy in the predicted total gating charge (sensibly higher than experimentally found), since the pore domain in real channels may limit the full movement of the $S_4$ segment, thus the gating charge being translocated. Consistent with this view is the observation that our modeled Q-V relationship could be fitted with a single Boltzmann relationship (Figure 10B), while two Boltzmann components are necessary to describe the Q-V relationship of Shaker channels (Bezanilla et al., 1994). We think that the second Boltzmann component needed in a real Shaker channel likely originates from the cooperative movement of the four $S_4$ segments suggested to represent the last step of the activation pathway, a feature not included in our model.

Finally our model was also able to predict the changes in kinetics and amount of gating currents observed in response to a depolarizing step preceded by different prepulse voltages, a process referred to as the Coole-Moore shift (Bezanilla et al., 1994). The details of these results are reported in Supplementary Figure 4.

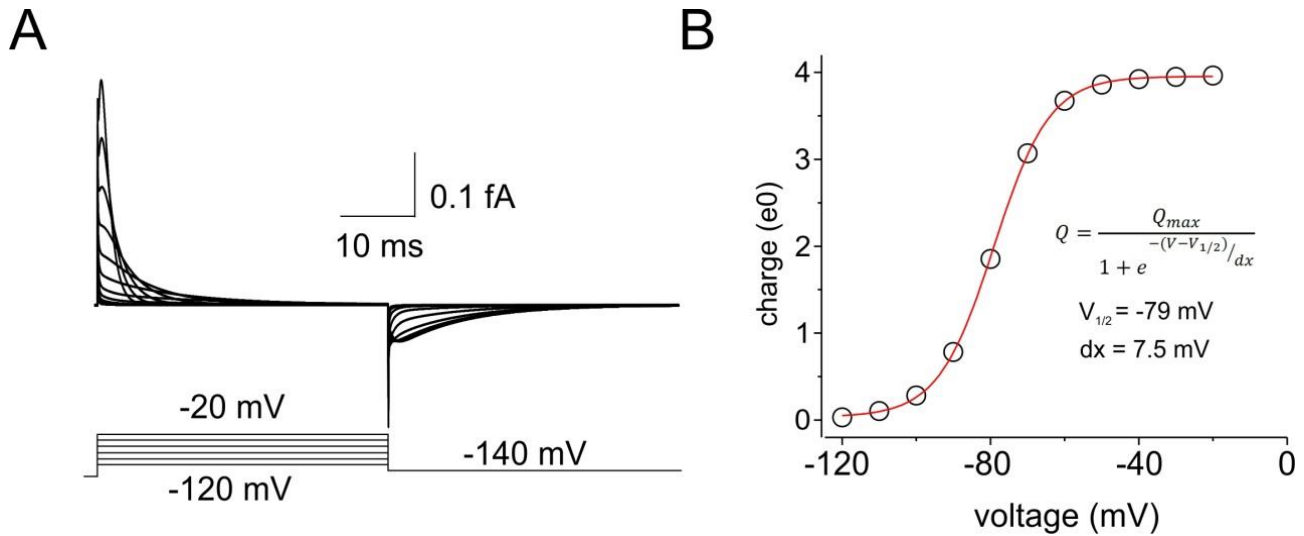

**Figure 10. A)** Family of simulated macroscopic gating currents evoked by voltage pulses from -120 to -20 mV (in steps of 10 mV) from a holding voltage of -140 mV. **B)** Plot of the gating charge, obtained as the integral of the macroscopic gating current, as a function of the applied voltage, for the simulations shown in panel A. The solid line represents the best fit of the simulated data with a single Boltzmann relationship, whose parameters are reported in the plot.

# Discussion

In this paper we have presented a new modeling approach to investigate the voltage-dependent gating mechanism of ion channels. Our model is capable to accurately reproduce the experimentally determined macroscopic gating current starting from the structural properties of the VSD, and to give an insight into the physical mechanism at the base of channel electromechanical transduction. The model treats the charged $S_4$ segment as a Brownian particle subjected to electro-diffusion, and assesses the electrical force acting on it by taking into account the externally applied voltage and the electrostatic voltage originating from the charged residues of the VSD and electrolyte ions. We verified the validity of the model by testing the conservation of the total current produced along the spatial domain knowing that the application of the Maxwell equations to models involving the movement of charges should produce a current that remains constant in space (Eisenberg et al., 2017, 2018) (cf Supplementary material).

Our model predicts a maximal $S_4$ displacement of 28.6 Å (as assessed from the mean distance between the $S_1$ and $S_5$ stable states of the segment, cf. Figure 2B). Although this displacement may appear too high, it must be considered that the $S_4$ segment is tilted by ~40-60° with respect to the membrane axis (Long et al., 2007), giving a vertical maximal displacement of the voltage sensor of 14-22 Å, values close to the 15-20 Å vertical displacement assessed in some experiments (Rhuta and MacKinnon, 2005).

The model predicts and explains most of the experimentally determined features of the macroscopic gating current. First, it predicts the existence of stable intermediate states of the $S_4$ segment, corresponding to the various positions the $S_4$ segment can take and the different number of gating charges exposed in the intracellular and extracellular vestibules. The presence of multiple closed states in Shaker K channels has long been suggested by studies of the gating and ionic currents, and incorporated into kinetic DMM of voltage-dependent gating (Bezanilla et al., 1994; Zagotta et al., 1994, Schoppa and Sigworth, 1998). More recently MD simulations (Delemotte et al., 2011) and mutagenesis experiments (Tao et al., 2010) have confirmed the presence of five substates assumed by the $S_4$ segment along its activation pathway. Our numerical simulations predict that these stable states are separated by energy barriers of 2-4 $k_BT$ at a membrane voltage close to the activation $V_{1/2}$. Notably, theoretical calculations performed using the Kramer's diffusion theory suggest that well defined markovian states (that may be well described by DMM) are delimited by barriers at least 4-5 $k_BT$ high (Sigg et al., 1999; Cooper et al., 1988). The energetic landscape resulting from our model is even more divergent from those described by DMM when the applied voltage is outside the range for channel activation. As we have shown in Figure 3, in this range a clear distinction of the energy profile into multiple stable states is somehow loosened, and the energetic landscape suggests a continuous drift motion of the $S_4$ segment along its activation pathway. These results suggest that channel gating properties not expected on the base of DMM may well appear in channel gating under particular conditions.

The model also predicts a very fast component in the macroscopic gating current at the beginning of a depolarizing and repolarizing step. In accordance with previous numerical simulations performed using continuous models of channel gating (Sigg et al., 1999), our model suggests that this fast component mainly originates from a redistribution of the $S_4$ segment population within the most extreme energy wells (fully activated or fully deactivated), due to the sudden change in the electrical force acting on the voltage sensors following a step change in the applied voltage. However, when the applied voltage change is performed within the range of channel activation, the occupancy of the adjacent stable state along the activation pathway may also participate to this fast component, likely because of the relatively low energy barrier separating the two stable states.

A third property of the voltage sensor dynamics well reproduced by our model is the peculiar shape of the macroscopic gating current time course for relatively high depolarizing steps, consisting of a rising phase followed by a slower decay. Using DMM, this behavior has been previously interpreted with initial transitions carrying less charge than later transitions along a chain of voltage sensor conformational states. This interpretation is not supported by currently available structural and functional data, which indicate that the various closed states of the channel correspond to different positions of the voltage sensor relative to the gating pore, while its charges interact sequentially with the GCTC (Tao et al., 2010). In this view, each transition from one closed state to the next should carry exactly the same – one – positive charge along the entire voltage drop (i.e. through the gating pore), and no gating transition carrying a different amount of charge could be postulated. Within this framework, our model was able to reproduce the initial rising phase of the macroscopic gating current by simply considering the discreteness of the charge density along the $S_4$ segment, due to the discrete nature of the gating charged residues, and the electrical field dropping almost entirely within the short gating pore, so that only the charge effectively present within this region is capable to catch the electrical field that is translated into the force acting on the $S_4$ segment. Given that the gating pore is relatively short as compared to the distance between the gating charges along the $S_4$ segment, the charge density residing within the gating pore as the $S_4$ segment moves along the activation pathway will be subjected to strong oscillations. As a result, at the beginning of a depolarizing step from, for instance, a highly negative holding voltage that concentrates virtually all the $S_4$ segments within only one state, this increasing charge passing inside the gating pore will cause a massive acceleration of all $S_4$ segments, that will in turn determine a rising macroscopic gating current. The oscillating behavior of the gating current disappears after the passage of the first gating charge through the gating pore, due to the fact that the thermal diffusion soon desynchronizes the $S_4$ segment populations.

Our model is also able to predict the complex time course of the OFF gating currents in response to repolarizations from different prepulse voltages, namely a monotonically decaying OFF gating current for prepulses below the channel activation $V_{1/2}$, and the presence of a second slower exponential component, or a rising phase preceding the gating current decay, for higher prepulse voltages. We found that this rising phase present in the OFF gating current has the same origin of the rising phase of the ON gating current at high depolarizing voltages, that we have described above. Although early mutagenesis experiments seemed to suggest that the initial slow rising phase of the OFF gating current represented an open state stabilization following the cooperative transition of the four channel subunits to the activated state (Batulan et al., 2010), more recent data have clarified that the slow onset and decay of the OFF gating current are preserved in the absence of pore opening, and thus represent an intrinsic property of the $S_4$ segment dynamics (Haddad and Blunck, 2011). Our model fully agrees with this view, as a rising phase of the OFF gating current can be predicted in the complete absence of a coupling of the $S_4$ voltage sensor with the channel pore.

Our simplified model cannot obviously account for several features of the gating mechanism, since provisions for them have not been included. First, being developed in one dimension, the model is not able to predict possible tilting and rotations of the $S_4$ segment during its trip along the activation pathway. Many experimental data suggest that while translating in the direction perpendicular to the membrane plane, the $S_4$ segment undergoes a 180° counterclockwise rotation, a movement that allows the positioning of the gating charges always close to the countercharges present in the $S_1$-$S_3$ segments of the VSD, thus maximizing their electrostatic interaction (Cha et al., 1999; Glauner et al., 1999; Grizel et al., 2014). This rotation is coherently predicted by MD simulations, that also suggest a tilting of the $S_4$ segment, changing from 60° to 35° the angle with the plane of the membrane during its activation (Pathak et al., 2007). Based on these data a future development of a 3D Brownian model will better describe the conformational changes of the $S_4$ segment during voltage-dependent gating. Second, our model only includes electrostatic interactions between the $S_4$ segment and the rest of the VSD. However other types of interactions such as Wan der Waals, $\pi$-cation, *etc*, may significantly contribute to the stability of the various kinetic substates of the VSD. Notably the contribution of non-electrostatic interactions may easily be introduced in the Langevin and Fokker-Planck equations as an additional external force contributing to the drift term. Finally, the Brownian model presented in this paper considers only one of the four VSDs contributing to the gating of an ion channel, and thus it cannot be used to explore cooperative interactions between the different channel subunits. Interestingly, inter-subunit cooperativity has long been postulated in Shaker K channels, on observing that their DMM kinetic schemes had to include a cooperative step preceding channel opening in order to explain the steep Q-V and G-V relationships at relatively depolarized voltages (Bezanilla et al., 1994; Zagotta et al., 1994; Schoppa and Sigworth, 1998). In addition mutagenesis experiments indicate that the N-terminal part of the $S_4$-$S_5$ linker of each Shaker K channel subunit interacts with the C-terminal part of the adjacent $S_6$ segment of another subunit during channel opening (Batulan et al., 2010), suggesting that a full understanding of the voltage-dependent gating should include inter-subunit interaction. In accordance with this view, we think that the failure of our model to predict a double Boltzmann as found in experimental Q-V relationships is due to the lack of a cooperative, inter-subunit conformational change. Our model may easily be expanded to include four interacting VSDs controlling a single channel gate. This expansion of the model would in addition confer the possibility to predict ionic currents in addition to gating currents, thus increasing the available experimental data that may be considered to understand the physics of the voltage-dependent gating.

# Materials and Methods

In our model the VSD was approximated to a hourglass-shaped geometrical structure consisting of a water inaccessible cylindrical gating pore, having a length of 0.2 nm and a diameter of 1 nm, flanked by internal and external water accessible vestibules having a length of 3.4 nm each and a conical shape opening with a half angle of 15° into two hemispherical subdomains of bath solution, both having radii of 1 μm (Figure 1A and B). The $S_4$ charge profile ($Z_{S4}$) was built by considering the six positive charges, whose mean distance between the charged atoms was determined from a 3D Shaker channel model (Figure 1A, see Supplementary material), and each giving rise to a charge profile normally distributed with a standard deviation of 0.1 nm (Supplementary Figure 1D). The fixed charged profile ($Z_F$) was similarly built by considering the vertical dimension of the $C_\alpha$ of all the charged residues of the S1-S3 part of the VSD (Figure 1A).

Ions were subjected to electro-diffusion governed by the following flux conservative equation:

$$\frac{dC_j(x,t)}{dt} = -\nabla F_j(x,t) \tag{1}$$

where $C_j(x,t)$ is the concentration of ion j, t is the time, $\nabla$ is the spatial gradient operator, and $F_j(x,t)$ is the flux (mole per second per unit area) of ion j, given by the Nernst-Planck equation:

$$F_j(x,t) = -D_j(x)\left[\nabla C_j(x,t) + \frac{z_j F}{RT} \nabla V(x,t)\right] \tag{2}$$

where $D_j(x)$ and $z_j$ are the diffusion coefficient profile and the valence of ion j, respectively, F, R and T have their usual meanings, and $V(x,t)$ is the electrical voltage profile.

The $S_4$ segment was assumed to move in one dimension as a Brownian particle, whose dynamics is governed by the following Langevin's equation

$$\dot{x}_{S4}(t) = F_{ex}(x_{S4},t)/\gamma + r(t) \tag{3}$$

Here $x_{S4}(t)$ represents the position of the voltage sensor (distance between the R2-R3 midpoint and the center of the gating pore), $m$ is the mass of the particle, $F_{ex}(x_{S4},t)$ is the external (electrical) force acting on the particle, and R(t) is a random force due to the collision of the fluid and the rest of the protein on the $S_4$ segment, which has a probability distribution with zero mean and second moment given by $<R(t)\,R(t')> = 2\,k_B\,T\,\delta(t-t')$ where $k_B$ is the Boltzmann constant and $\delta$ is the delta function. $\gamma$, the friction coefficient of the $S_4$ voltage-sensor, was settled to 4*10$^{-6}$ Kg/s based on the comparison between experimental and predicted macroscopic gating currents.

The electrical voltage profile V(x) was assessed from the net charge density profile $\rho(x)$, using the following Poisson's equation

$$\varepsilon_0 \left[\frac{d}{dx}\left(\varepsilon(x)\frac{dV(x)}{dx}\right) + \varepsilon(x)\frac{dV(x)}{dx}\frac{d\ln A(x)}{dx}\right] = -\rho(x) \tag{4}$$

Where $\varepsilon_0 = 8.854\cdot 10^{-12}$ F·m$^{-1}$ is the vacuum permittivity, $\varepsilon(x)$ is the position dependent dielectric coefficient, $V(x)$ is the electric potential and $A(x)$ is the position dependent surface. More details on the model are available in Supplemental Material. Programs were written in C and are available upon request. Figures were prepared using Origin v4.0 and CorelDraw v18.1.

# Acknowledgments

The authors thanks Dr Wolfgang Nonner and Dr Bob Eisenberg for critically reading the Ms and giving us useful suggestions.

# Supplemental Material

# The model

***Assessment of the Shaker K channel 3D structure by homology modeling***

In our model the geometrical and electrostatic properties of the VSD have to be derived from the 3D structure of the channel under study (*cf.* below). While this structural information is presently available for the Kv1.2 and Kv1.2/Kv2.1 channel chimera, most of the functional data (i.e. gating and ionic current measurements) have been obtained from the Shaker K channel (*cf* Introduction of the paper). Although these two K channels have a high degree of homology and are thus expected to have a very similar 3D structure (Long et al., 2007), differences in charged residues in relevant positions of the VSD may well change the details of their gating behavior. We thus proceeded to generate a 3D structure of the Shaker K channel by homology modeling, using the SWISS MODEL environment (Biasini et al., 2014; Arnold et al., 2006; Guex et al., 2009) and the Kv1.2/Kv2.1 chimera as a template.

Homology modeling, also known as comparative protein structure modeling, is a computational approach to build 3D structural models for proteins using experimental structures of related protein family members as templates. It generates the structural coordinates of the model based on the mapping between the target residues and the corresponding amino acids of the structural template. Regions of the protein for which no template information is available (i.e. insertions and deletions in loop regions) are built from libraries of backbone fragments or by *de novo* reconstruction by constrained dynamics. Local suboptimal geometry of the models obtained (i.e. distorted bonds, angles, and too close atomic contacts) are finally regularized by limited MD. The Shaker channel primary sequence (accession P08510.3) was first aligned to the Kv1.2/Kv2.1 chimera using the PDBviewer software v4.10 (see Supplementary Figure 1F), and then the two aligned sequences, together with the 3D coordinates of the Kv1.2/Kv2.1 chimera, were sent to the SWISS-MODEL web site in order to find the 3D structure for the Shaker K channels. As shown in Supplementary Figure 1A, the 3D structure model for the Shaker channel appears very similar to that experimentally found for the Kv1.2/Kv2.1 channel chimera, as expected from the very high similarity of the two protein sequences. Notwithstanding, several residues predicted to reside close to the $S_4$ segment appear to be differently charged in the VDS of the two channels (see Supplementary Figure 1A, where the negative and positive residues of the two structures are colored in red and yellow, respectively, and the gating charges on the $S_4$ segment in magenta). This is especially evident in their external vestibules, with the Shaker VSD charge density profile displaying a strongly negative charge density peak not present in the chimera channel. Diverse dynamics of the $S_4$ segment in the two channels, due to the marked difference in the electrostatic voltage profile along their gating pores and vestibules, can thus be envisaged. This peculiar feature is expected to stabilize the $S_4$ segment in its activated position, in accordance with a Q-V relationship shift to more hyperpolarized voltages found for the Shaker K channel as compared to the Kv1.2 channel, which has a fixed charge profile virtually identical to the Kv1.2/Kv2.1 chimera (Bezanilla et al., 1994; Ishida et al., 2015).

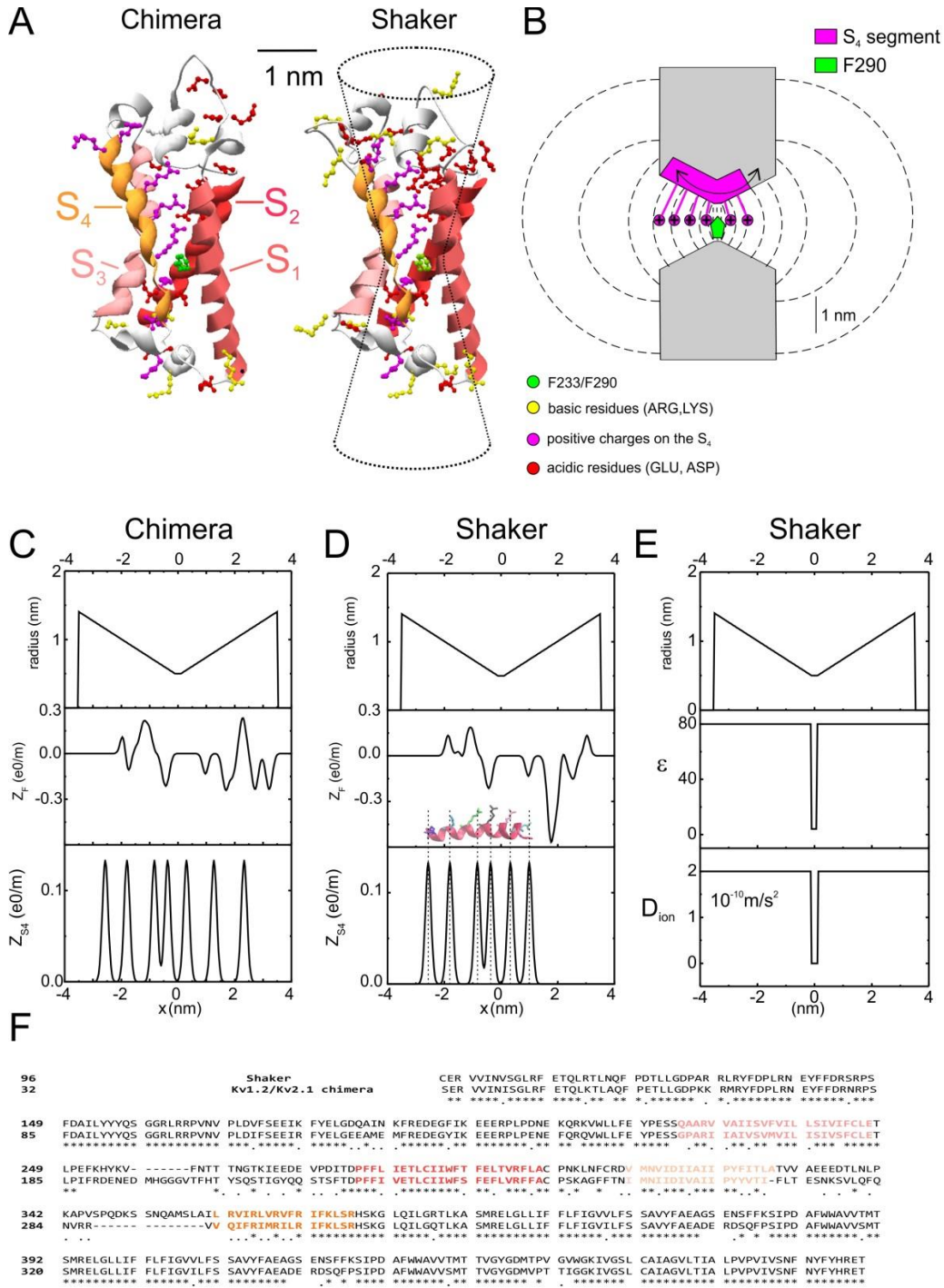

**Supplementary Figure 1. A)** Representation of one VSD from the Kv1.2/Kv2.1 chimera (structure 2R9R, left) in the activated state, and a model of the Shaker 3D structure obtained by homology modeling using the 2R9R structure as template (right). In both structures gating charges on the $S_4$ segment are in magenta, while negative and positive residues located in the remaining parts of the VSDs are in red and yellow, respectively. The residues F233 (in the chimera) and F290 (in Shaker) are in green. The hourglass-shaped drawing superimposed to the Shaker structure represents the geometry used in our model to delimit the gating pore and vestibules. The superimposition shows that the choice of 15° as half angle aperture approximates quite well the shape of the vestibules. **B)** Schematics showing the geometry of the VSD assumed in our model. The $S_4$ segment containing the 6 gating charges was assumed to move perpendicular to the membrane through the gating pore (0.2 nm long) and the extracellular and intracellular vestibules (each 3.4 nm long, and opening with a half angle of 15°). The dashed lines represent some of the surfaces delimiting the volume elements considered in our numerical simulations (see text for details). **C)** and **D)** Profiles of the gating pore radius, the fixed charge density located in the S1-

S3 region of the VSD ($Z_F$), and the charge density on the $S_4$ segment ($Z_{S4}$), for the Kv1.2/Kv2.1 chimera (left) and the Shaker model structure (right). For symmetrical reasons, x=0 was assumed to coincide with the center of the gating pore, where the F233/F290 residue is assumed to be located. *Inset:* 3D structure of the $S_4$ segment region of the Shaker channel model, with the R1-R6 charged residues explicitly shown in liquor ice representation. Notice the reasonable correspondence between the peaks of the $Z_{S4}$ profile and the charged atoms of the R1-R6 residues. **E)** Profiles of the gating pore radius, the relative dielectric constant ($\varepsilon$) and the electrolyte ion diffusion coefficients ($D_{ion}$) as a function of the spatial coordinate considered in our model (x). **F)** Sequence alignment for the Kv1.2/Kv2.1 chimera and the Shaker K channels. * indicates conserved residues, while . indicates residues with similar polarity.

### *Structure of the voltage sensor domain*

In our model the VSD was approximated by a hourglass shaped geometrical structure consisting of a water inaccessible cylindrical gating pore, having a length of 0.2 nm and a diameter of 1 nm, flanked by internal and external water accessible vestibules having a length of 3.4 nm each and a conical shape opening with a half angle of 15° into two hemispherical subdomains of bath solution, both having radii of 1 $\mu$m (Supplementary Figure 1A and B). As shown in Supplementary Figure 1A, this geometrical shape well adapted to that of the vestibules inferred from the 3D crystal structure, and a vestibule length of 3.4 nm ensures the $S_4$ segment to remain within the vestibules for all its allowed positions. The water inaccessible gating pore was located at the level of the F290 residue, proposed to separate the internal and external vestibules of the VSD. This symmetrical geometry allows the formulation of the model in one spatial dimension consisting in a main axis perpendicular to the membrane and passing through the gating pore. In the numerical simulation the main axis was divided into subdomains of constant step-size within the VSD, and a step-size increasing geometrically going outwards in the two bath solutions. Using this subdivision the surfaces separating adjacent sub-volumes were circles inside the gating pore, spherical caps in the vestibules, and hemispheres in the baths, each one contacting perpendicularly the channel wall (dashed lines in Supplementary Figure 1B). As emphasized in Supplementary Figure 1B, the $S_4$ segment does not occupy space in either vestibules, since it contributes to form the vestibule walls together with the other parts of the voltage sensor domain (S1-S3), in accordance with the available crystal structure showing that the extracellular vestibule is formed by a departure of the S3-S4 segments from the S1-S2 (Long et al., 2007). With regard to the charges on the $S_4$ segment, we explicitly consider them by letting them contribute to the charge density of the volume grids together with the charges carried by the other parts of the voltage sensor domain (fixed charges) and by freely moving ions. The $S_4$ charge profile ($Z_{S4}$, expressed in $e_0$ units) was built by considering six positive charges, whose mean distance was determined from the position of the charged atoms in the crystal structure, and each giving rise to a charge profile normally distributed with a standard deviation of 0.1 nm (Supplementary Figure 1D). Notice that the charged atoms at the top of the arginine and lysine lateral chains seems to maximized their distance in the crystal structure, so that the peak to peak distance in the $Z_{S4}$ profile is sensibly larger than the 4.5-6.0 Å expected for C$\alpha$ carbons of the corresponding residues in an $\alpha$ (or $\alpha$ 3-10) helix (inset to Supplementary Figure 1). The fixed charged profile ($Z_F$, expressed in $e_0$ units) was similarly built by considering all the charges of the S1-S3 part of the VSD (Supplementary Figure 1D). In our model the $S_4$ segment was assumed to be a rigid body, its position being represented by the variable $x_{S4}$, expressing the distance of the midpoint between R2 and R3 from the center of the gating pore. The position of the $S_4$ segment in Supplementary Figure 1B and D corresponds to $x_{S4}$=0, and during the simulation it was allowed to move through the gating pore and vestibules up to a maximal displacement $x_{S4}$ of ±1.8 nm, a movement that enables all the gating charges (R1 to K5) to reach the GCTC.

### *Ion electro-diffusion*

We assumed that the intracellular and extracellular faces of the VSD are bathed by ionic solutions containing 140 mM of positively and negatively charged monovalent ions, that can freely move in the baths and vestibules of the VSD with a diffusion constant of $2 \cdot 10^{-10}$ m/s$^2$, but cannot enter in the gating pore. Due to water and ions inaccessibility, the gating pore was assumed to have a relative dielectric constant ($\varepsilon$=4) much lower than in the bathing solution ($\varepsilon$=80; cf. Supplementary Figure 1E). Ions were subjected to electro-diffusion governed by the following flux conservative equation:

$$\frac{dC_j(x,t)}{dt} = -\nabla F_j(x,t) \quad (1)$$

where $C_j(x,t)$ is the concentration of ion j, t is the time, $\nabla$ is the spatial gradient operator, and $F_j(x,t)$ is the flux (mole per second per unit area) of ion j, given by the Nernst-Planck equation:

$$F_j(x,t) = -D_j(x)\left[\nabla C_j(x,t) + \frac{z_j F}{RT}\nabla V(x,t)\right] \quad (2)$$

where $D_j(x)$ and $z_j$ are the diffusion coefficient profile and the valence of ion j, respectively, F, R and T have their usual meanings, and $V(x,t)$ is the electrical voltage profile.

*Movement of the $S_4$ segment*

The $S_4$ segment was assumed to move in one dimension as a Brownian particle, whose dynamics is governed by the following Langevin's equation

$$m\,\ddot{x_{S4}}(t) = F_{ex}(x_{S4},t) - \gamma \dot{x_{S4}}(t) + R(t) \quad (3)$$

Here $x_{S4}(t)$ represents the position of the voltage sensor (distance between the R2-R3 midpoint and the center of the gating pore), $m$ is the mass of the particle, $F_{ex}(x_{S4},t)$ is the external (electrical) force acting on the particle, and R(t) is a random force due to the collision of the fluid and the rest of the protein on the $S_4$ segment, which has a probability distribution with zero mean and second moment given by:

$$<R(t)\,R(t')> = 2\,k_B\,T\,\delta(t-t') \quad (4)$$

where $k_B$ is the Boltzmann constant and $\delta$ is the delta function. $\gamma$, the friction coefficient of the $S_4$ voltage-sensor, was settled to 4*10$^{-6}$ Kg/s based on the comparison between experimental and predicted macroscopic gating currents.

In the very high friction limit the acceleration of the particle may be assumed to be zero, hence the Langevin equation reduces to:

$$\dot{x_{S4}}(t) = F_{ex}(x_{S4},t)/\gamma + r(t) \quad (5)$$

that may be written in the form of the following stochastic differential equation:

$$dx_{S4}(t) = \left(F_{ex}(x_{S4},t)/\gamma\right)dt + \sqrt{\frac{2\,k_B\,T\,dt}{\gamma}}\,\phi(t) \quad (6)$$

where $\phi(t)$ represents a normally distributed random variable with zero mean and unitary variance. Based on eqn. (6), the position of the particle may be found at each time-step dt as $x_{S4} = x_{S4}^{old} + dx_{S4}$, where $x_{S4}^{old}$ represents the starting position of the particle. The particle, as already stated, was allowed to freely move in the range $x_{S4} = \pm 1.8\,nm$, by imposing elastic boundary conditions.

In our model the external force acting on the $S_4$ segment, $F_{ex}(x_{S4},t)$, is represented by the electrical force due to the voltage gradient acting on the gating charges:

$$F_{ex}(x_{S4},t) = -e_0 \int Z_{mS4}(\epsilon)\left(\frac{dV(\epsilon,t)}{d\epsilon}\right)d\epsilon \quad (7)$$

Where $Z_{mS4}(\epsilon) = Z_{S4}(\epsilon + x_{S4})$ represents the gating charge spatial distribution when the S$_4$ segment is positioned at x$_{S4}$.

## *Assessment of the electrical voltage*

The electrical voltage profile *V(x)* was assessed from the net charge density profile $\rho(x)$, using the following Poisson's equation

$$\varepsilon_0 \left[\frac{d}{dx}\left(\varepsilon(x)\frac{dV(x)}{dx}\right) + \varepsilon(x)\frac{dV(x)}{dx}\frac{d\ln A(x)}{dx}\right] = -\rho(x) \tag{8}$$

Where $\varepsilon_0$ = 8.854·10$^{-12}$ F·m$^{-1}$ is the vacuum permittivity, $\varepsilon(x)$ is the position dependent dielectric coefficient, $V(x)$ is the electric potential and $A(x)$ is the position dependent surface. The charge density profile was assessed by including the gating charges, the fixed charges present in the remaining part of the VSD (Z$_F$, Supplementary Figure 1D), and the ions in solution:

$$\rho(x) = \frac{e_0(Z_F(x) + Z_{mS4}(x))}{A(x)\,dx} + e_0 \sum_j z_j C_j(x) \tag{9}$$

## *Simulation of the dynamics of a single S$_4$ segment*

In our simulation of a single S$_4$ segment eqns. (1), (6), and (8) were discretized in space and time using a variable step-size (dx=0.33 A in the VSD and an increasing step-size dx$_i$=2* dx$_{i-1}$ going far in the baths) and a time-step of 6 ns. Both eqns. (1) and (8) were solved using a fully implicit method and an appropriate algorithm for tridiagonal matrix equations, while the stochastic differential eqn. (6) was solved using a normally distributed random number generator from Press et al. (1992).

## *The Fokker-Planck equation*

The dynamics of the S$_4$ segment may also be described in terms of the time evolution of the probability density function profile, given by the following Fokker-Planck (FP) equation:

$$\frac{df_{S4}(x,t)}{dt} = -\frac{\partial(\mu(x) f_{S4}(x,t))}{\partial x} + \frac{k_B T}{\gamma}\frac{\partial^2 f_{S4}(x,t)}{\partial x^2} \tag{10}$$

Here $f_{S4}$ represents the probability density function for the position of the S$_4$ segment and $\mu(x)$ is $F_{ex}(x_{S4}, t)/\gamma$. This partial differential equation was solved with elastic boundary conditions at $x = \pm 1.8\ nm$ in order to set the allowed movement of the particle to 3.6 nm. The elastic boundary condition was imposed by setting the particle flux equal to zero at the boundaries:

$$Flux = \left(\mu f - \frac{k_B T}{\gamma}\frac{\delta f(x,t)}{\delta x}\right)_{x=\pm 1.8\ nm} = 0 \tag{11}$$

The solution of the FP equation relies on the possibility of finding a good description of the position-dependent drift velocity $\mu$. This is not a trivial issue, since $\mu$ will actually depend also on the relaxation time of electrolyte ions around the S$_4$ segment, that in turn will affect the electrical voltage and the external force acting on the segment. For this reason, if one considers the full model that includes electrolyte ions dynamics (eqn. 1), $\mu$ will become also a function of time, making non trivial the numerical solution of the FP equation. To solve this problem, in the solution of the FP equation we assumed steady-state for the dynamics of the electrolyte ions:

$$\frac{dC_j}{dt} = 0 \tag{12}$$

The validity of this approximation, which is based on the finding that ions relax on a timescale much faster than the movement of the S$_4$ segment, is fully demonstrated below (cf. paragraph 'Validation of the steady-state approximation for ion dynamics in the solution of the Fokker-Planck equation'). This approximation allowed to find a steady-state solution for the ion concentration and electrical voltage profiles for each allowed position of the S$_4$ segment within the volumes grid used in the numerical simulation. As a consequence, a $\mu$ spatial profile could easily be found with eqn. (7) and used during the time dependent simulation of the voltage sensor.

Once determined, the dynamics of the probability density function $f_{S4}(x,t)$ was used to estimate the macroscopic gating currents, to be compared with those obtained experimentally. However, since the solution of the FP equation required the steady-state approximation for the electrolyte ions dynamics (*cf.* above), the gating current could not be directly obtained from eqn. (2), but was assessed by analyzing the net charge changes in the left (or alternatively in the right) bath, according to the following equation (Horng et al., 2017):

$$I_g(t) = \frac{d \int_{-1.8\,nm}^{1.8nm} f_{S4}(x,t) \left( dQ_{bath}(x)/dx \right) dx}{dt} \tag{13}$$

Where $Q_{bath}(x_{S4}) = e_0 \int_{left\,bath} A(x) \left( \sum_j z_j C_j \right) dx = e_0 \int_{righ\,bath} A(x) \left( \sum_j z_j C_j \right) dx$ represents the net charge found in the left (or right) bath when the position of the S$_4$ segment is x$_{S4}$, and the integration covers the whole left (right) bath.

**Validation of the steady-state approximation for ion dynamics in the solution of the Fokker-Planck equation**
In order to numerically find the probability density function of the S$_4$ segment's position it is necessary to make an approximation in our model, consisting in the assumption that the electrolyte ions equilibrates instantaneously (cf. above). This steady-state approximation is very reasonable, since electrolyte ions move at a rate much faster than the movement of the S$_4$ segment. We however verified its validity by comparing the output of the full model used in the simulation of the single S$_4$ segment dynamics with that of a reduced model containing the described approximation.

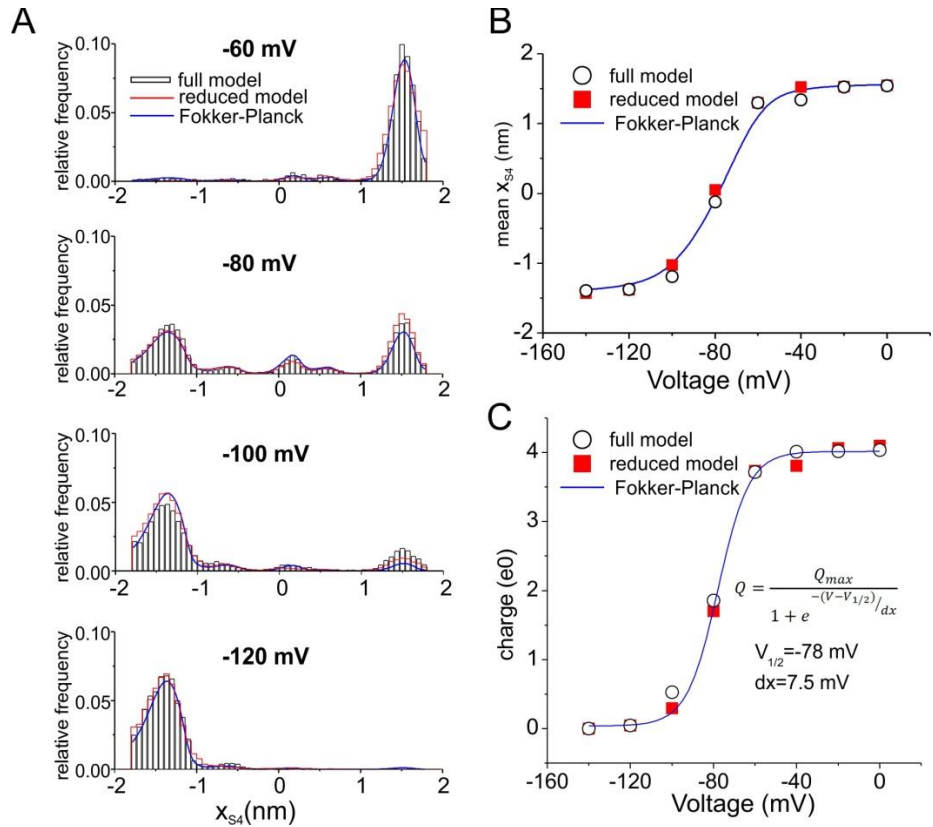

**Supplementary Figure 2. A)** Amplitude histograms of the $S_4$ segment position $x_{S4}$, obtained from 100 ms simulations at the indicated voltages. The black columns are simulation obtained with the full model, also shown in Figure 2 of the Ms. The red lines represent amplitude histograms obtained by running stochastic simulations of the reduced model, assuming instantaneous steady-state for the electrolyte ion concentrations. The blue lines represent the probability density function of the $S_4$ segment position, found by solving the FP equation up to equilibrium, at the four different applied voltages. **B)** Plot of the mean $x_{S4}$ as a function of the applied voltage, assessed using the full stochastic (circles, also reported in Figure 2C of the Ms), the reduced stochastic (red squares) and the FP (blue line) models. **C)** Plot of the mean charge vs the applied voltage, obtained by integrating the microscopic gating current over 50 ms long simulations at different applied voltages. The simulations used are the same reported in Figure 5 on the Ms. black and red symbols refer to simulations performed using the full or the reduced model, respectively. The solid blue line represents the fit of the full model data with a Boltzmann relationship, with best fit parameters indicated in the Figure.

In the Supplementary Figure 2A we compared the amplitude histograms of the S4 segment positions built from simulations obtained with the full model – that is, the amplitude histograms already shown in Figure 2B of the paper – with the amplitude histograms obtained from simulations done using the reduced model, represented by the superimposed red lines. The blue lines in the same Figure represent instead the predicted probability density function of the $S_4$ segment position, obtained by solving the FP equation, thus also including the above mentioned approximation. It is evident that the differences between the curves derived from the full and the reduced models are within the variability originating from the stochastic nature of the simulations, thus validating the approximation. In Supplementary Figure 2B we plot the mean $S_4$ position vs voltage assessed for the full model (data already shown in Figure 2C of the Ms) and compared it with that obtained by using either single particle simulations obtained from a reduced model or by directly solving the FP equation. Also in this case a full agreement was obtained. Finally, we also compared the full and reduced models in predicting the behavior of the gating currents originating from the movement of a single $S_4$ segment. As shown in Supplementary Figure 2C, the voltage dependence of the mean charge displacement assessed from the time integral of the gating currents results very similar between the two models, again validating the steady-state approximation for electrolyte ions electro-diffusion.

**Comparison between experimental and modeled time courses of the ON and OFF gating currents.**
Gating currents of the Shaker channel have been studied in detail in the attempt to obtain information on the mechanism leading to the voltage-sensor conformational change in response to membrane depolarization. At the macroscopic level, gating currents recorded from a population of channels show complex properties depending on the amplitude of the depolarization, decaying mono-exponentially for small depolarizations, bi-exponentially at intermediate voltages, and showing a rising phase followed by a fast exponential decay for relatively high depolarizations. Notable, a very fast decaying component with few tens microseconds time constant was found when macroscopic gating currents are recorded at high bandwidth (Sigg et al., 2003), and interpreted with the voltage sensor rapidly moving in a low energy well right upon the depolarization begins (Sigg et al., 1999). Also the gating currents recorded on repolarization from a series of depolarizing steps (OFF gating current) display very complex properties, decaying fast and mono-exponentially when the prepulse voltage is relatively low, bi-exponentially for intermediate prepulse voltages, and with the appearance of a rising phase when evoked from relatively high prepulse voltages (Bezanilla et al., 1994; Stefani et al., 1994).

Notable, all this features of the macroscopic gating current are qualitatively reproduced by our model (Supplementary Figure 3). More specifically, the ON gating currents, besides a very fast mono-exponential decaying component with a time constant of ~100 μs present for all depolarizing steps, displayed: i) an additional decaying exponential component with time constant of ~1 ms for relatively small depolarizations (-120 and -110 mV); ii) two additional decaying exponential components with time constants of ~1.5 and 15 ms for slightly higher depolarizations (-80 mV); iii) a plateau or rising phase followed by a mono-exponential decay with time constant of ~9 ms for depolarizations at which most of the channels reach the fully activated state (-60 to -20 mV). As for the OFF gating current, in addition to a very fast mono-exponential decaying component with time constant of ~120 – 20 μs (depending on the prepulse voltage) we obtained: i) an additional decaying component with time constant of 0.8 ms for very small depolarizing prepulses (-120 mV); ii) two additional decaying exponential components with time constant of ~0.7 and ~8 ms for slightly more depolarized prepulse voltages (-100 and -80 mV);

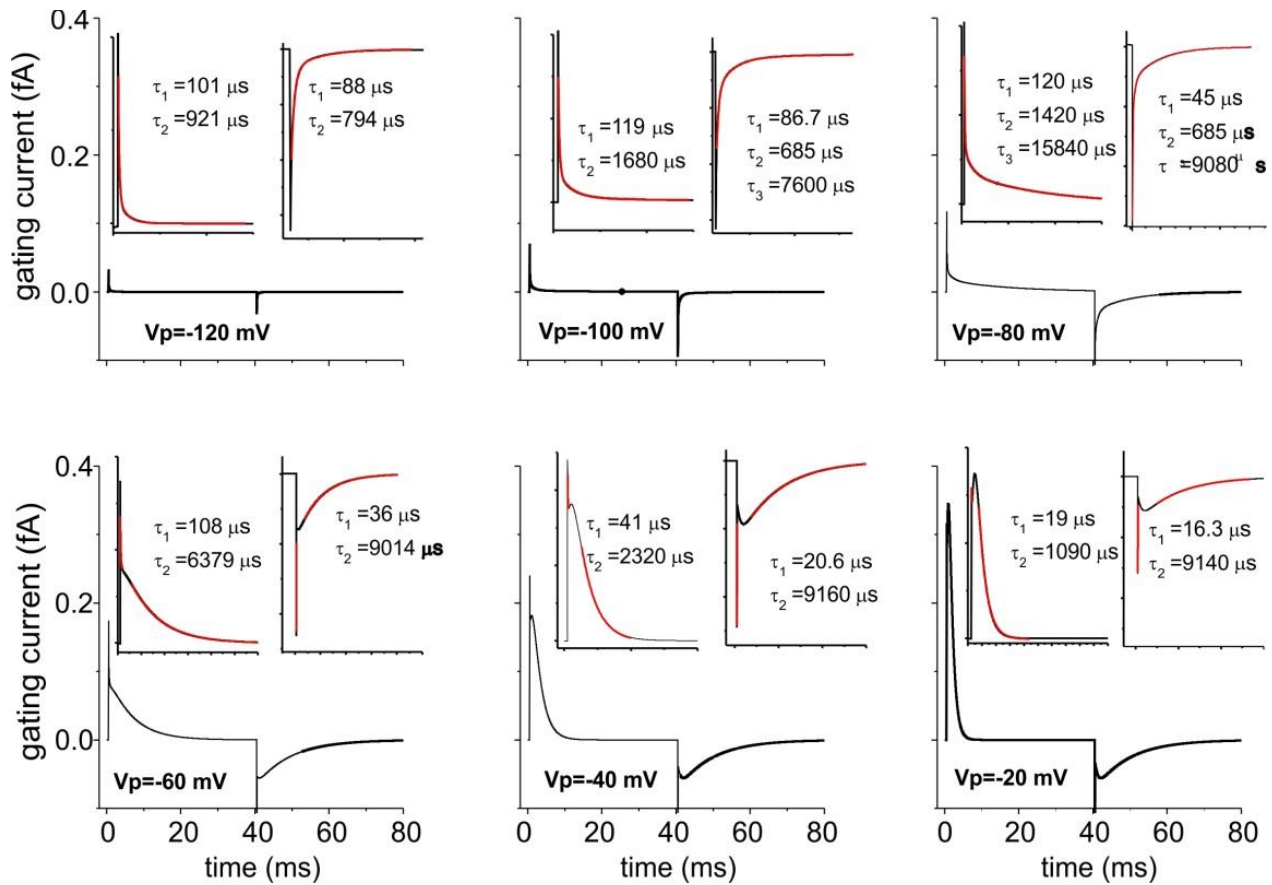

**Supplementary Figure 3.** Each plot displays in isolation a simulated macroscopic gating current obtained in response to a single depolarization (at the indicated applied voltage). The insets are expansions of the ON and OFF gating currents, with superimposed multi-exponential fits performed with the minimum number of exponentials needed to obtain a satisfactory fit. Best fit time constants are indicated in the Figure.

iii) a plateau or rising phase followed by a mono-exponential decay with time constant of ~9 ms for prepulse voltages at which most of the channels reach the fully activated state (-60 to -20 mV). All these features of the ON and OFF gating currents qualitatively match with those experimentally found for the Shaker K channel (Bezanilla et al., 1994).

*The Cole-Moore effect on gating currents*
Many different types of Kv channels, including Shaker K channels, display an activation rate strongly dependent on the magnitude of the prepulse voltage, an effect often referred to as Cole-Moore shift (Cole and Moore, 1960). This is usually taken as a demonstration of the presence of a linear sequence of voltage-dependent closed state transitions that the channel must undergo before opening. Since a change in the prepulse voltage alters the occupancy of the different closed states at rest, also the time needed to reach the open state in response to a depolarizing step will change, with higher rates obtained with more depolarized voltages, where closed states proximal to the open state became occupied. A similar Cole-Moore shift can also be observed in Shaker gating currents, where strong depolarizations from relatively hyperpolarized voltages produce gating currents with a clear rising phase and a subsequent delayed decay, while identical depolarizations from less hyperpolarized voltages lead to the disappearance of the rising phase and the consequent anticipation of the decaying phase of the gating current (Bezanilla et al., 1994). In order to verify whether a Cole-Moore shift is reproduced by our model of Shaker K channel gating current, we simulated a two-pulse protocol and looked at the gating current elicited at 0 mV after applying 5 ms long prepulses at different membrane voltages. As shown in Supplementary Figure 4A, our model correctly predicts that, as the prepulse voltage becomes more depolarized, the rising phase of the gating current at 0 mV gradually disappears, and the decaying phase is significantly anticipated.

In order to understand the mechanism at the origin of the Cole-Moore shift of the macroscopic gating currents, in Supplementary Figure 4B we plotted the time course of the mean position of the $S_4$ segment, the amount of gating change inside the gating pore, and the force acting on the $S_4$ segment. With relatively hyperpolarized prepulse voltages the $S_4$ segment mostly occupies state 1, with a probability distribution that predicts a very small charge inside the gating pore (~0.05 e0). With such a small charge within the electrical field, the force pushing the $S_4$ segment extracellularly upon stepping to 0 mV will be proportionally small. However, as the $S_4$ segments will coherently move extracellularly, their charges inside the gating pore will gradually increase, leading to an increase in the applied force and velocity of the segments. The combination of the time dependent increase of the gating charge traversing the pore and the velocity of the $S_4$ segment accounts for the rising phase of the gating current at the beginning of the depolarization. As the prepulse voltage becomes more depolarized, the gating charge stably residing in the gating pore at the beginning of the test pulse will progressively increase (Supplementary Figure 4B), and so will the force acting on the $S_4$ segment at the beginning of the test pulse, and the $S_4$ segment acceleration. This will cause the disappearance of the rising phase in the macroscopic gating current.

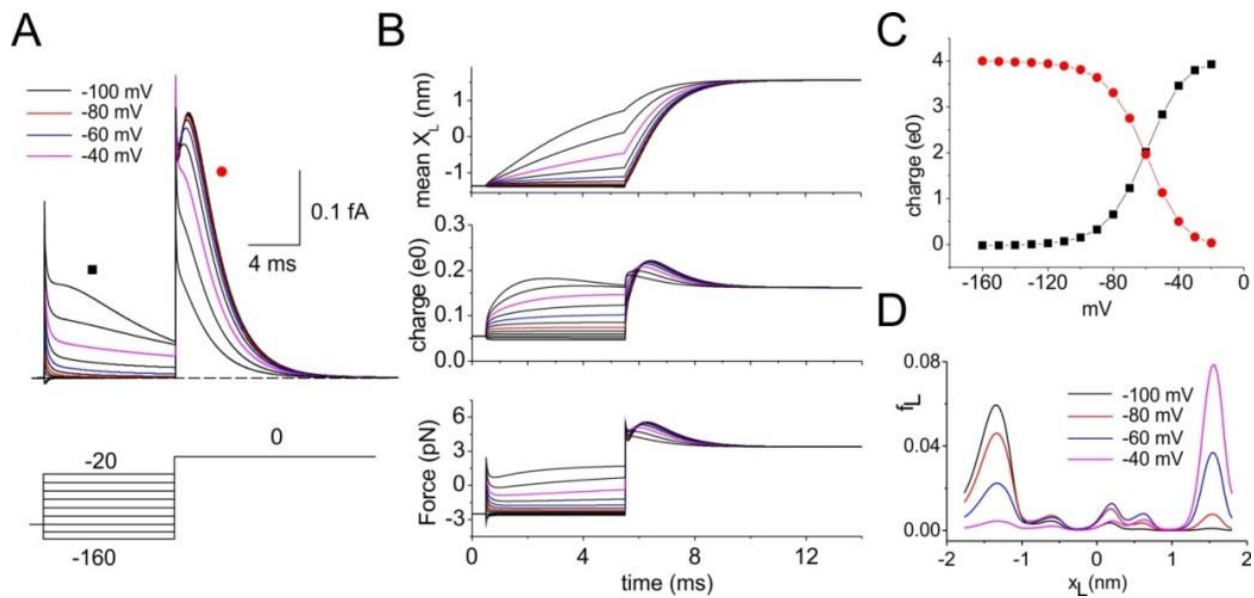

**Supplementary Figure 4. A)** Family of simulated macroscopic gating currents evoked by a voltage protocol consisting of a test pulse to 0 mV preceded by 5 ms long prepulses ranging from -160 to -20 mV (in steps of 10 mV). **B)** Time courses of the mean $S_4$ segment position, charge inside the gating pore, and force applied on the $S_4$ segment, obtained for the simulations of panel A. **C)** Plot of the gating charge, obtained as the integral of the macroscopic gating current during the test pulse (red symbols) and during the prepulse (black symbols), as a function of the prepulse voltage, for the simulations shown in panel A. **D)** Profiles of the probability density function of the $S_4$ segment at the end of the 5 ms prepulse, from the simulations shown in panel A.

*Validation of the model: conservation of the total current*

We validated our model by verifying that the total current produced was conserved along the spatial domain. It has been recently shown that the application of the Maxwell equations to models involving the movement of charges gives rise to a very simple rule that applies independently to the details of the model and the time scale considered: the current produced by moving masses, when summed up to a displacement current, proportional to the temporal changes in the electric field, results in a total current that should remain constant in space (Eisenberg et al., 2017, 2018).

Although our model for the macroscopic gating current considers a population of $S_4$ segments, we first consider only one $S_4$ segment inside its voltage sensor domain and surrounded by K and Cl ions in the baths and vestibules. For this system a current conservation can be written for each type of moving charge of the system

$$\frac{d\rho_j(x,t)}{dt} = -\frac{d(i_j(x,t)/A(x))}{dx} \tag{S1}$$

where $\rho_j(x,t)$ is the charge density (charge per unit volume) of species j (in our model either monovalent anion and cation, or the charged $S_4$ segment), t is the time, and $i_j$ is the current produced by species j (charge per unit time), and $A(x)$ is the surface normal to the particle flux. Summing up the current conservation equations for all species we obtain:

$$\frac{d\rho(x,t)}{dt} = -\frac{di(x,t)/A(x)}{dx} \tag{S2}$$

With $\rho(x,t) = \sum \rho_{ions}(x,t) + \rho_{S4}(x,t)$ being the total moving charge density, and $i(x,t) = i_{ions}(x,t) + i_{S4}(x,t)$ being the particle current.

In our model $i_{ions}(x,t)$ is assessed on the assumption that ions cannot pass through the gating pore, and by applying charge conservation. More specifically:

$$i_{ions}(x,t) = \frac{d}{dt}\int_x^{x_{pl}} A(x)\, F\left(\sum_{j=0}^{nions-1} c_j(x,t)z_j\right) dx = -\frac{d}{dt}\int_{x_{pr}}^x A(x)\, F\left(\sum_{j=0}^{nions-1} c_j(x,t)z_j\right) dx \tag{S3}$$

Where $x_{pl}$ and $x_{pr}$ are the left and right extremes of the gating pore, $F$ is the Faraday constant, and $z_j$ is the valence of ion j. From similar considerations, $i_{S4}(x,t)$ can be assessed as

$$i_{S4}(x,t) = \frac{d}{dt}\int_x^L e_0\, z_{mS4}(x,t)\, dx = -\frac{d}{dt}\int_0^x e_0\, z_{mS4}(x,t)\, dx \tag{S4}$$

Where $z_{S4}(x,t)$ is the valence density profile of the $S_4$ segment.

Finally, in our model all the charges contribute to shape the electric field E in accordance with the Gauss law, that in the differential and mono-dimensional form reads

$$\varepsilon_0 \frac{d[A(x)\, \varepsilon(x)\, E(x, x_4, t)]}{dx} = \rho^*(x,t) \tag{S5}$$

Where $\rho^*(x,t) = \sum \rho_{ions}(x,t) + \rho_{S4}(x,t) + \rho_F$, with $\rho_F$ being the time- and position-independent fixed charge, and $E(x, x_4, t)$ is the electric field, for which we have explicitly indicated the dependence on the spatial dimension, time, and position of the voltage sensor $x_{S4}$. Taking the time derivative of eqn. (S5) we obtain

$$\varepsilon_0 \frac{d}{dx}[A(x)\,\varepsilon(x)\frac{dE(x,x_4,t)}{dt}] = \frac{d\rho^*(x,t)}{dt} = \frac{d\rho(x,t)}{dt} \tag{S6}$$

And combining eqns (S2) and (S6) we obtain

$$\frac{d}{dx}[i_{tot}(x,t)] = 0 \tag{S7}$$

which shows the conservation of the total current defined as:

$$i_{tot}(x,t) = i_{ions}(x,t) + i_{S4}(x,t) + i_{dipls}(x,t) \tag{S8}$$

where we have introduced the displacement current

$$I_{dipls}(x,t) = A(x)\,\varepsilon_0\,\varepsilon(x)\frac{dE(x,x_4,t)}{dt} \tag{S9}$$

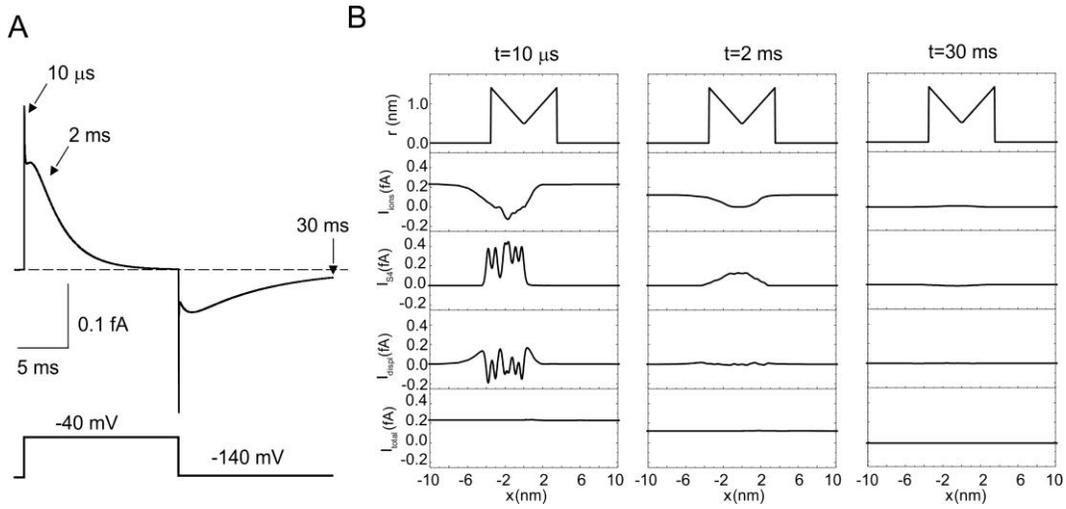

**Supplementary Figure 5. A)** Simulated macroscopic gating current evoked by a voltage pulse from -140 to -40 mV (protocol indicated below). **B)** Plot of the radius of the vestibules and gating pore, and of the ionic ($I_{ions}$), $S_4$ segment ($I_{S4}$), and displacement ($I_{displ}$) currents as a function of the spatial dimension (x), at three different times from the beginning of the depolarization (indicated).

In our model we actually consider a population of $S_4$ segments, distributed in the allowed positions $x_{S4}$ in accordance with the density function $f_{S4}(x_4, t)$, assessed by solving the Fokker Planck equation. In order to find a conservation equation to apply to the mean macroscopic gating current, we integrate eqn. (S6) for all possible positions of the $S_4$ segment, weighting with the density function $f_{S4}(x_4, t)$.

$$\int_{-x_L/2}^{x_L/2} \frac{d}{dx}[A(x)\,\varepsilon_0\varepsilon(x)\frac{dE(x,x_4,t)}{dt} + i(x,t)]\,f_{S4}(x_{S4})\,dx_{S4} = 0 \tag{S10}$$

Where $\pm x_L/2$ represent the extreme positions allowed to the $S_4$ segment. Rearranging

$$\frac{d}{dx}[I_{dipls}(x,t) + I_{ions}(x,t) + I_{S4}(x,t)] = 0 \tag{S11}$$

Where

$$I_{dipls}(x,t) = \frac{d}{dt}\int_{-x_L/2}^{x_L/2}[A(x)\,\varepsilon_0\varepsilon(x)E(x,x_4,t)]\,f_{S4}(x_{S4})\,dx_{S4} \tag{S12}$$

$$I_{ions}(x,t) = \int_{-x_L/2}^{x_L/2} i(x,t)\,f_{S4}(x_{S4})\,dx_{S4} = \frac{d}{dt}\int_{-x_L/2}^{x_L/2}\left[\int_x^{x_{pl}} A(x)\,F\left(\sum_{j=0}^{nions-1} c_j(x,t)z_j\right)\right]f_{S4}(x_{S4})\,dx_{S4} \tag{S13}$$

$$I_{S4}(x,t) = \frac{d}{dt}\int_{-x_L/2}^{x_L/2}\left[\int_x^L e_0\,z_{S4}(x,x_{S4})\,dx\right]f_{S4}(x_{S4})\,dx_{S4} \tag{S14}$$

Supplementary Figure 5 shows the total current profile, together with the three contributing currents (ionic, $S_4$ segment, and displacement currents), assessed using eqns (S12), (S13) and (S14) at three different times (10μs, 2 ms, and 30 ms ) from the beginning of a depolarizing pulse from -140 to -40 mV. As expected. the conservation of the total current is respected.